\documentclass[a4paper,twocolumn]{article}

\usepackage{amssymb,amsfonts,amsmath,amstext,amsgen,amsopn,amsxtra,indentfirst,graphicx,color,ulem}

\usepackage{geometry}
\begin{document}

\title{Closed-formed ab initio solutions of geometric albedos \\ and reflected light phase curves of exoplanets} 
\date{}
\maketitle
\vspace{-0.5in}
\noindent
\author{\scriptsize \textbf{Kevin Heng$^{1,2,3\dagger}$, Brett M. Morris$^{1,4}$, Daniel Kitzmann$^1$}}\\
\author{\scriptsize $^1$ University of Bern, Center for Space and Habitability, Gesellschaftsstrasse 6, CH-3012, Bern, Switzerland. Emails: kevin.heng@unibe.ch, daniel.kitzmann@unibe.ch}\\
\author{\scriptsize $^2$ University of Warwick, Department of Physics, Astronomy \& Astrophysics Group, Coventry CV4 7AL, United Kingdom. Email: Kevin.Heng@warwick.ac.uk}\\
\author{\scriptsize $^3$ Ludwig Maximilian University, University Observatory Munich, Scheinerstrasse 1, Munich D-81679, Germany. Email: Kevin.Heng@physik.lmu.de}\\
\author{\scriptsize $^4$ University of Bern, Physics Institute, Division of Space Research \& Planetary Sciences, Sidlerstrasse 5, CH-3012, Bern, Switzerland. Email: brett.morris@unibe.ch}\\
\author{\scriptsize $^\dagger$ Corresponding author}

\textbf{Studying the albedos of the planets and moons of the Solar System dates back at least a century \cite{bond1861,russell1916,horak50,vau64}.  Of particular interest is the relationship between the albedo measured at superior conjunction, known as the ``geometric albedo", and the albedo considered over all orbital phase angles, known as the ``spherical albedo" \cite{russell1916,sobolev,seager}.  Determining the relationship between the geometric and spherical albedos usually involves complex numerical calculations \cite{horak65,dy74,hh89,marley99,sudarsky00} and closed-form solutions are restricted to simple reflection laws \cite{van74,madhu12}.  Here we report the discovery of closed-form solutions for the geometric albedo and integral phase function, which apply to any law of reflection that only depends on the scattering angle.  The shape of a reflected light phase curve, quantified by the integral phase function, and the secondary eclipse depth, quantified by the geometric albedo, may now be self-consistently inverted to retrieve globally averaged physical parameters. Fully Bayesian phase curve inversions for reflectance maps and simultaneous light curve detrending may now be performed due to the efficiency of computation. Demonstrating these innovations for the hot Jupiter Kepler-7b, we infer a geometric albedo of $0.25^{+0.01}_{-0.02}$, a phase integral of $1.77 \pm0.07$, a spherical albedo of $0.44^{+0.02}_{-0.03}$ and a scattering asymmetry factor of $0.07^{+0.12}_{-0.11}$.}

The Bond albedo $A_{\rm B}$ is the fraction of starlight that is reflected by a celestial body---either by its atmosphere or surface---over all viewing angles and wavelengths of light.  For example, the Bond albedo of the Earth is about 0.3.  The spherical albedo $A_s$ is considered over all angles, but only at one wavelength of light.  The geometric albedo $A_g$ is the albedo measured when an exoplanet is at superior conjunction---when the star is between the observer on Earth and the exoplanet.  This occurs at an orbital phase angle $\alpha$ of zero.  The phase integral $q$ relates the spherical and geometric albedos \cite{russell1916,sobolev},
\begin{equation}
A_s = q A_g.
\end{equation}
The phase integral is notoriously difficult to evaluate \cite{seager} and requires knowledge of the integral phase function $\Psi$ \cite{russell1916,sobolev},
\begin{equation}
q = 2 \int^{\pi}_0 \Psi ~\sin\alpha ~d\alpha.
\label{eq:phase_integral}
\end{equation}
An exact, closed-form solution for $\Psi$ exists for a Lambertian sphere \cite{russell1916,sobolev,seager}, which is a perfectly reflective sphere that is equally bright regardless of viewing direction.

\begin{figure}
\begin{center}
\includegraphics[width=\columnwidth]{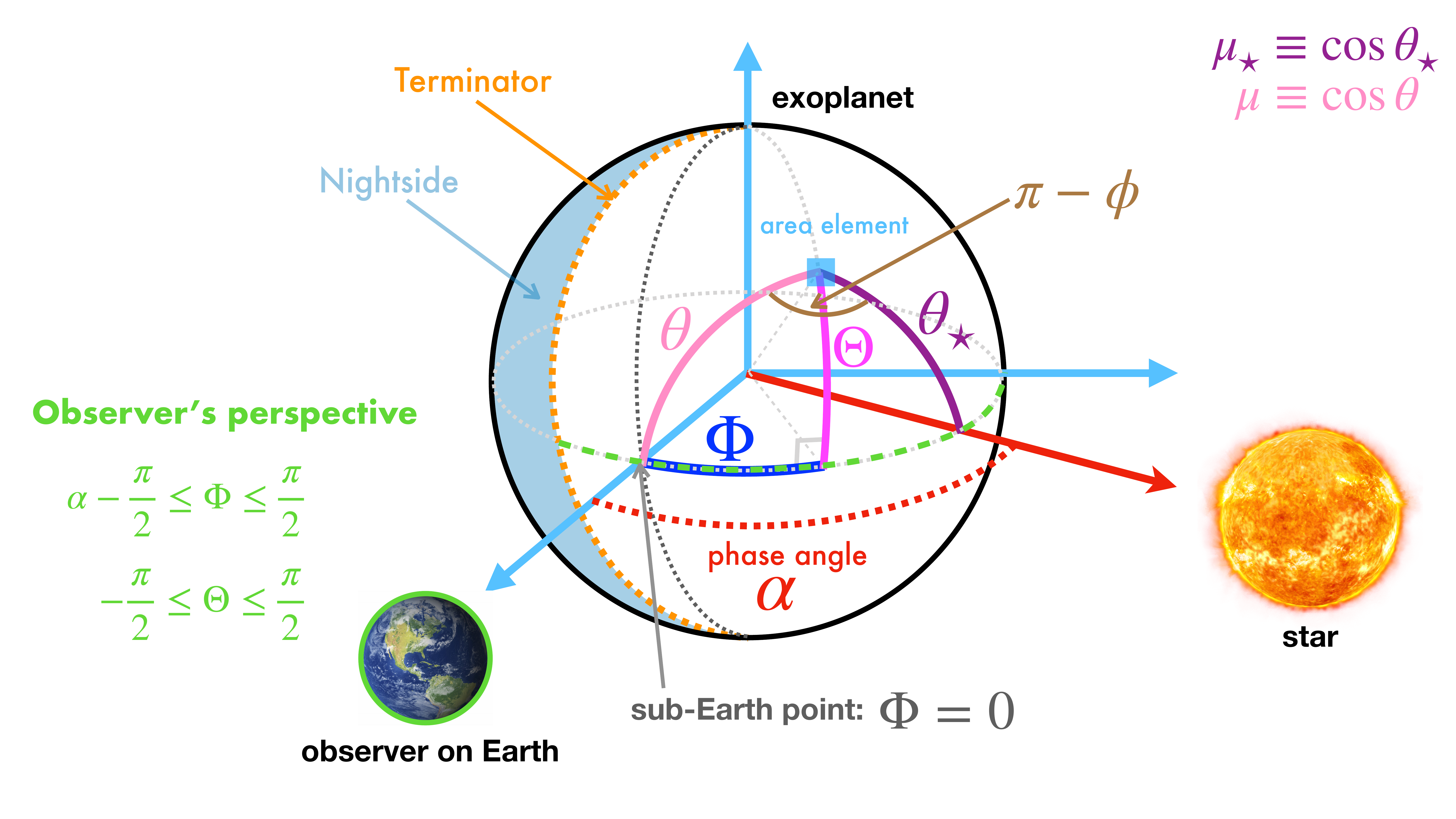}
\end{center}
\vspace{-0.2in}
\caption{Schematic describing the geometry of the system. An observer on Earth views an exoplanet with observer-centric latitude $\Theta$ and longitude $\Phi$.  At each point in its orbit, the exoplanet resides at a phase angle $\alpha$.  Starlight impinges upon the exoplanet at an angle $\theta_\star$ from its zenith.  An infinitesimal area element on the exoplanet may be described by the local coordinate system $(\theta,\phi)$. Trios of these angles form triangles on the surface of the sphere, and may be related by the spherical law of cosines \cite{horak50,sobolev,horak65}.}
\vspace{-0.1in}
\label{fig:geometry}
\end{figure}

Insight is gained by considering the geometry of the problem, which may be formulated in two different coordinate systems \cite{horak50,sobolev,horak65} (Figure \ref{fig:geometry}).  The observer-centric coordinate system defines a latitude $\Theta$ and longitude $\Phi$, such that the location on the exoplanet closest to Earth (the sub-Earth point) always sits at $\Phi=0$.  The position of the exoplanet in its orbit around the star is defined by a phase angle $\alpha$, such that the exoplanet is illuminated only across $\alpha - \pi/2 \le \Phi \le \pi/2$ as seen by the observer.  The local coordinate system defines a polar angle $\theta$ and an azimuthal angle $\phi$ for each point on the exoplanet.  Starlight impinges upon each point at a zenith angle $\theta_\star$.  The observer- and local coordinate systems are generally not aligned in the same plane (Figure \ref{fig:geometry}).  In principle, calculations may be performed in either coordinate system and all six angles are mathematically related (see Methods).  The coordinate systems are agnostic about the orbital inclination of the exoplanet.

All of the quantities of interest ($A_g$, $A_s$, $\Psi$, $q$) fundamentally involve the reflection coefficient \cite{sobolev,seager,dy74,hh89,van74},
\begin{equation}
\rho = \frac{I_0}{I_\star \mu_\star},
\end{equation}
which relates the intensities of incident starlight $I_\star$ and reflected starlight $I_0$.  We have defined $\mu_\star = \cos\theta_\star$.  By definition, a Lambertian sphere has $\rho=1$ \cite{sobolev}.  The geometric albedo may be evaluated in the local coordinate system \cite{sobolev,dy74},
\begin{equation}
A_g = \frac{\int^1_0 \rho_0 I_\star \mu^2 ~d\mu}{\int^1_0 I_\star \mu ~d\mu} = 2 \int^1_0 \rho_0 \mu^2 ~d\mu,
\label{eq:geometric_albedo}
\end{equation}
where we have defined $\mu = \cos\theta$.  The reflection coefficient evaluated at zero phase angle ($\alpha=0$) is represented by $\rho_0$.  The integral phase function may be written as $\Psi = F/F_0$ and  is evaluated in the observer-centric coordinate system using the dimensionless Sobolev flux \cite{sobolev}
\begin{equation}
F = \int^{\pi/2}_{\alpha-\pi/2} \int^{\pi/2}_0 \rho \mu \mu_\star ~\cos\Theta ~d\Theta ~d\Phi,
\label{eq:flux_sobolev}
\end{equation}
where $F_0$ is $F$ evaluated at $\alpha=0$.  Alternatively, the geometric albedo may be evaluated in the local coordinate system as $A_g = 2F_0/\pi$ \cite{sobolev}.

The reflection coefficient is derived from the radiative transfer equation (see Methods).  If the approximation is made that single scattering may be described by any reflection law, but multiple scattering of light occurs isotropically \cite{hapke81}, then the reflection coefficient of a semi-infinite atmosphere is
\begin{equation}
\rho = \frac{\omega}{4} \frac{1}{\mu_\star + \mu} \left( P_\star - 1 + H H_\star \right),
\end{equation}
where $H(\mu)$ is Chandrasekhar's H-function \cite{chandra} and $H_\star=H(\mu_\star)$ (see Methods).  The single-scattering albedo $\omega$ is the fraction of light reflected during a single scattering event.  The reflection law or scattering phase function $P_\star$ relates the incident direction of collimated starlight to the emergent direction of scattered starlight (see Methods).  It is usually assumed to depend only on the difference in angle between these directions, which is known as the scattering angle $\beta$ \cite{pierrehumbert}.  For example, isotropic and Rayleigh scattering are described by $P_\star=1$ and $P_\star = \frac{3}{4} ( 1 + \cos^2\beta )$, respectively.  The scattering and phase angles are trivially related,
\begin{equation}
\cos\beta = - \cos\alpha.
\end{equation}
When strong forward scattering occurs (scattering asymmetry factor of $g>0.5$), the assumption of isotropic multiple scattering breaks down (see Methods).
 
A single reflection law is assumed to apply throughout the atmosphere.  More realistically, the abundance, composition and sizes of aerosols vary in three dimensions within an atmosphere.  Using a single reflection law implies that only globally averaged values of these physical parameters are retrieved.  For example, the globally averaged spherical albedo of Jupiter may be retrieved across wavelength (see Methods).

At zero phase angle, one simply has $\cos\beta=-1$.  Correspondingly, the scattering phase function at $\alpha=0$ is evaluated at $\mu=\mu_\star$ and $\phi=\pi$ (see Methods) and is represented by $P_0$.  For example, $P_0 = 3/2$ for Rayleigh scattering.  Since $P_0$ is a \textit{number} and not a function, the geometric albedo for \textit{any} scattering law is straightforwardly derived (see Methods),
\begin{equation}
A_g = \frac{\omega}{8} \left( P_0 - 1 \right) + \frac{\epsilon}{2} + \frac{\epsilon^2}{6} + \frac{\epsilon^3}{24},
\end{equation}
where $\epsilon = (1-\gamma)/(1+\gamma)$ is the bihemispherical reflectance \cite{hapke81} and $\gamma = \sqrt{1-\omega}$.  The bihemispherical reflectance is typically small unless one approaches the pure reflection limit ($\omega=1$).   The term involving $P_0$ accounts for the single scattering of starlight, while the other terms account for isotropic multiple scattering.  For $\omega=1$ and isotropic single scattering ($P_0=1$), one obtains $A_g = 17/24$, which is slightly higher than the $A_g=2/3$ value of the Lambertian sphere by 1/24.

\begin{figure*}
\begin{center}
\vspace{-0.4in}
\includegraphics[width=\columnwidth]{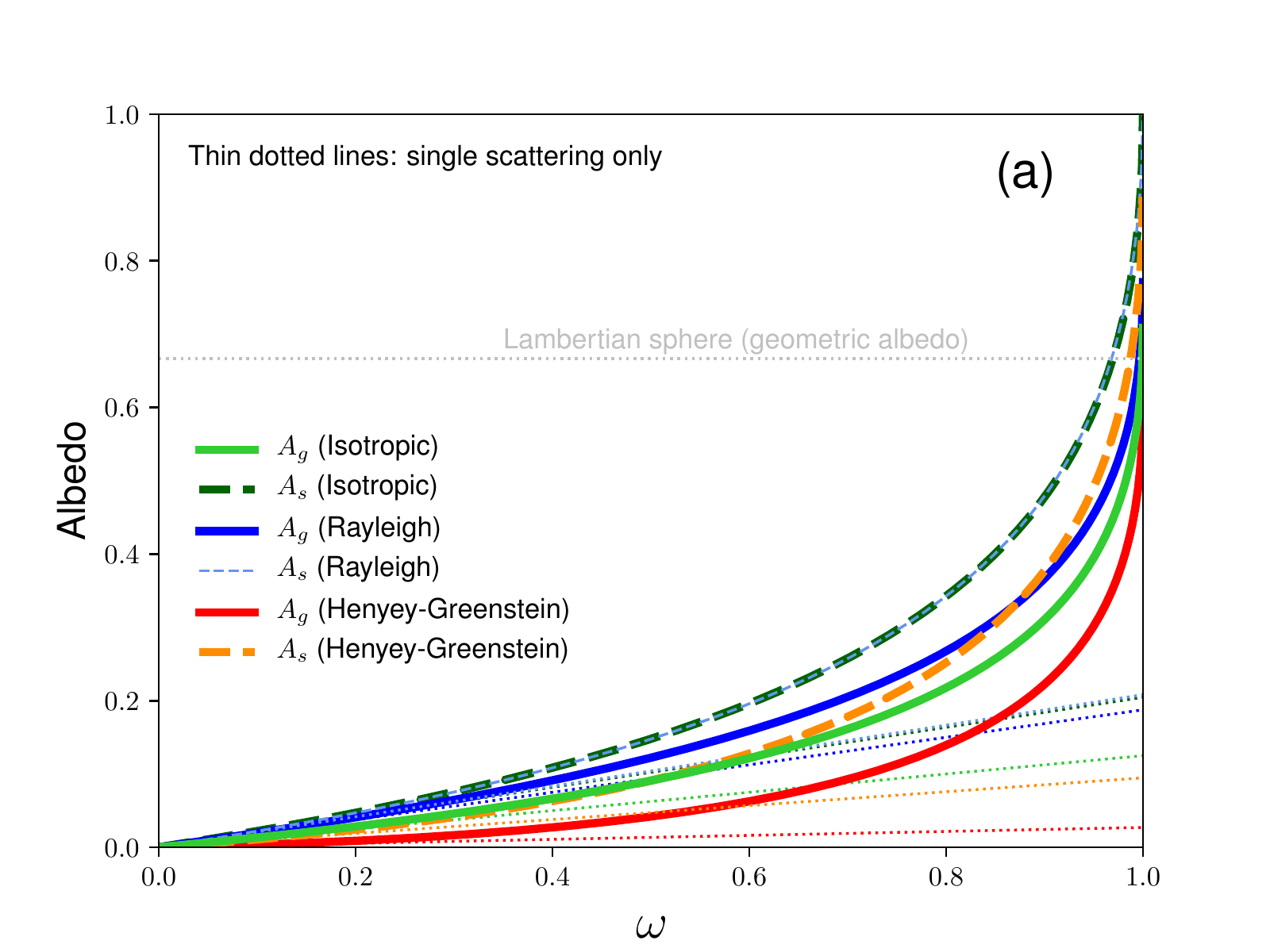}
\includegraphics[width=\columnwidth]{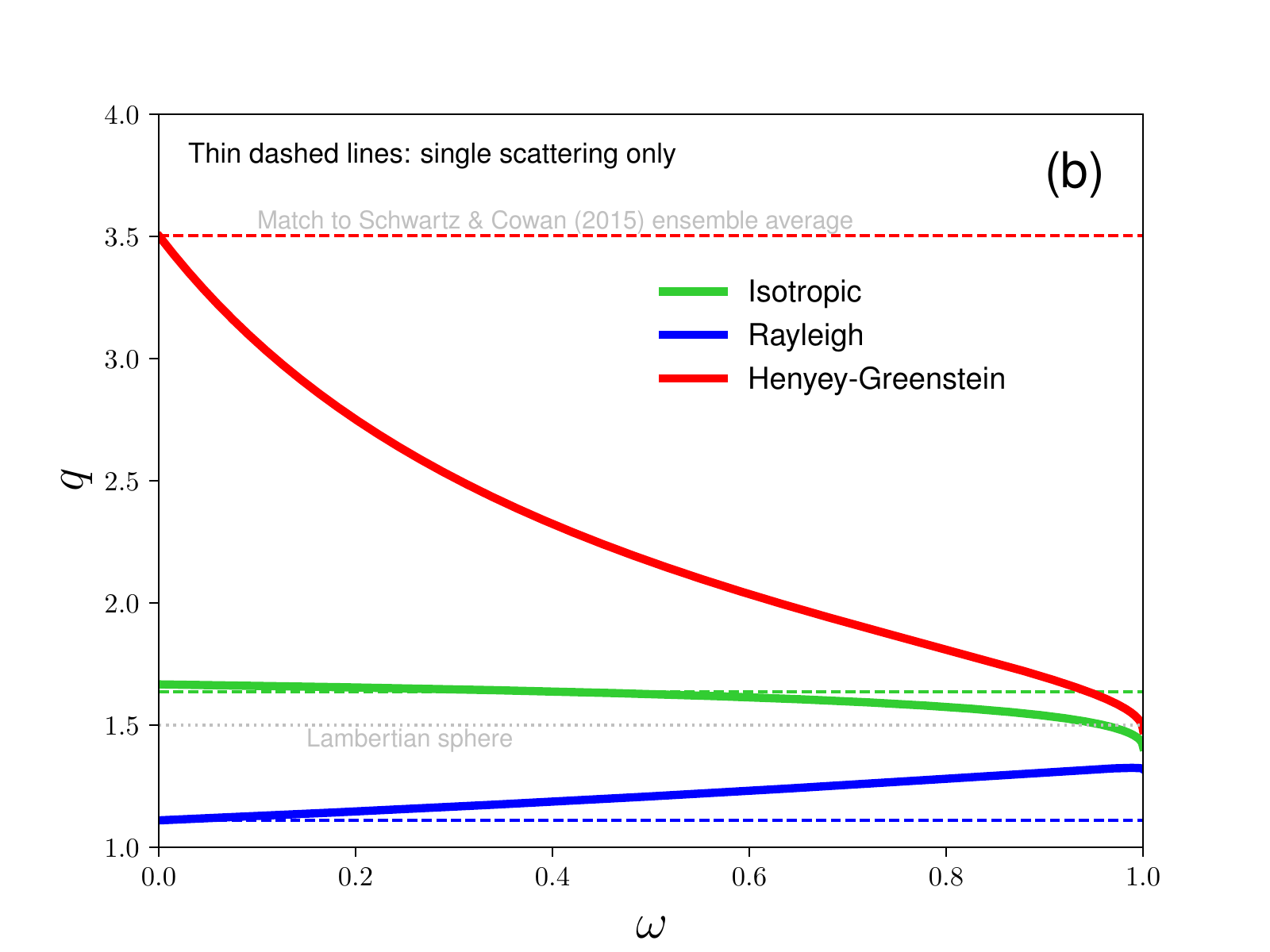}
\end{center}
\caption{(a) Geometric and spherical albedos and (b) phase integral for the isotropic, Lambertian, Rayleigh and Henyey-Greenstein reflection laws. Thick curves include multiple scattering, while the thin lines exclude it. For illustration, the scattering asymmetry factor of the Henyey-Greenstein reflection law is chosen to be $g=0.508$ such that the phase integral (single scattering only) has a value of 3.5, which matches the ensemble averaged value of a population of hot Jupiters \cite{sc15}. The Lambertian sphere has geometric and spherical albedos of 2/3 and 1, respectively.}
\vspace{0.1in}
\label{fig:albedos_q}
\end{figure*}

The integral phase function $\Psi$ may be derived using the stated reflection coefficient.  In the observer-centric coordinate system, the two-dimensional integral over $\Theta$ and $\Phi$ may be evaluated as two independent, one-dimensional integrals.  In the local coordinate system, $\alpha$ has a complicated relationship with $\theta$, $\theta_\star$ and $\phi$.  In the observer-centric coordinate system, $\alpha$ is instead the third independent angle, which allows $P_\star(\alpha)$ to be taken out of the integrals involving $\Theta$ and $\Phi$.  These insights render the derivation of $\Psi$ analytically tractable (see Methods).

Physically, $\Psi$ consists of three components.  The first component $\Psi_{\rm S}$ exhibits single-scattering behaviour.  The second component $\Psi_{\rm L}$ behaves like a Lambertian sphere \cite{hapke81}.  The third component $\Psi_{\rm C}$ is a correction term that is non-negligible only when $\omega \sim \epsilon \sim 1$ (strong reflection).  Our derivation yields
\begin{equation}
\Psi = \frac{12 \rho_{\rm S} \Psi_{\rm S} + 16 \rho_{\rm L} \Psi_{\rm L} + 9 \rho_{\rm C} \Psi_{\rm C}}{12 \rho_{\rm S0} + 16 \rho_{\rm L} + 6 \rho_{\rm C}},
\end{equation}
where the various coefficients are
\begin{equation}
\begin{split}
\rho_{\rm S} &= P_\star - 1 + \frac{1}{4} \left( 1 + \epsilon \right)^2 \left( 2 - \epsilon \right)^2, \\
\rho_{\rm S0} &= P_0 - 1 + \frac{1}{4} \left( 1 + \epsilon \right)^2 \left( 2 - \epsilon \right)^2, \\
\rho_{\rm L} &= \frac{\epsilon}{2} \left( 1 + \epsilon \right)^2 \left( 2 - \epsilon \right), \\
\rho_{\rm C} &= \epsilon^2 \left( 1 + \epsilon \right)^2. \\
\label{eq:rho_coefficients}
\end{split}
\end{equation}
The three components of the integral phase function are
\begin{equation}
\begin{split}
\Psi_{\rm S} &= 1 - \frac{1}{2} \left[ \cos{\left(\frac{\alpha}{2}\right)} - \sec{\left(\frac{\alpha}{2}\right)} \right] \Psi_0, \\
\Psi_{\rm L} &= \frac{1}{\pi} \left[ \sin{\alpha} + \left( \pi - \alpha \right) \cos{\alpha} \right], \\
\Psi_{\rm C} &= -1 + \frac{5}{3} \cos^2{\left(\frac{\alpha}{2}\right)}  - \frac{1}{2} \tan{\left(\frac{\alpha}{2}\right)} \sin^3{\left(\frac{\alpha}{2}\right)}\Psi_0, \\
\Psi_0 &= \ln{\left[ \frac{\left(1 + \alpha_-\right) \left(\alpha_+ - 1 \right)}{\left(1 + \alpha_+ \right) \left(1 - \alpha_- \right)} \right]},
\end{split}
\label{eq:Psi}
\end{equation}
and we have defined $\alpha_\pm = \sin(\alpha/2) \pm \cos(\alpha/2)$.  For $-\pi \le \alpha \le 0$, one needs to replace $\alpha$ by $\vert \alpha \vert$.

While these formulae may appear long and unwieldly, they constitute a closed-form solution of the integral phase function for \textit{any} reflection law $P_\star(\alpha)$ that includes both single and (isotropic) multiple scattering.  The phase integral $q$ may be straightforwardly obtained by a numerical integration of $\Psi$ over the phase angle.  The spherical albedo is $A_s = q A_g$ (see Methods for an alternative way of computing $A_s$).  Figure \ref{fig:albedos_q} shows calculations of $A_g$, $A_s$ and $q$ for the isotropic, Rayleigh and Henyey-Greenstein reflection laws.  Multiple scattering is dominant in determining the values of the geometric and spherical albedos.  For example, if only isotropic single scattering is considered, the maximum values of the geometric and spherical albedos are 1/8 and $2(1-\ln2)/3 \approx 0.2$, respectively.  If only single scattering is considered, the phase integral is independent of the single-scattering albedo $\omega$.  With multiple scattering included, $q$ has a non-trivial dependence on $\omega$, especially for the Henyey-Greenstein reflection law.  We validated our calculations of $A_g$ and $A_s$ against those from previous studies \cite{dy74,van74,madhu12} for various reflection laws (see Methods).

\begin{figure*}
\begin{center}
\vspace{-0.2in}
\includegraphics[width=\columnwidth]{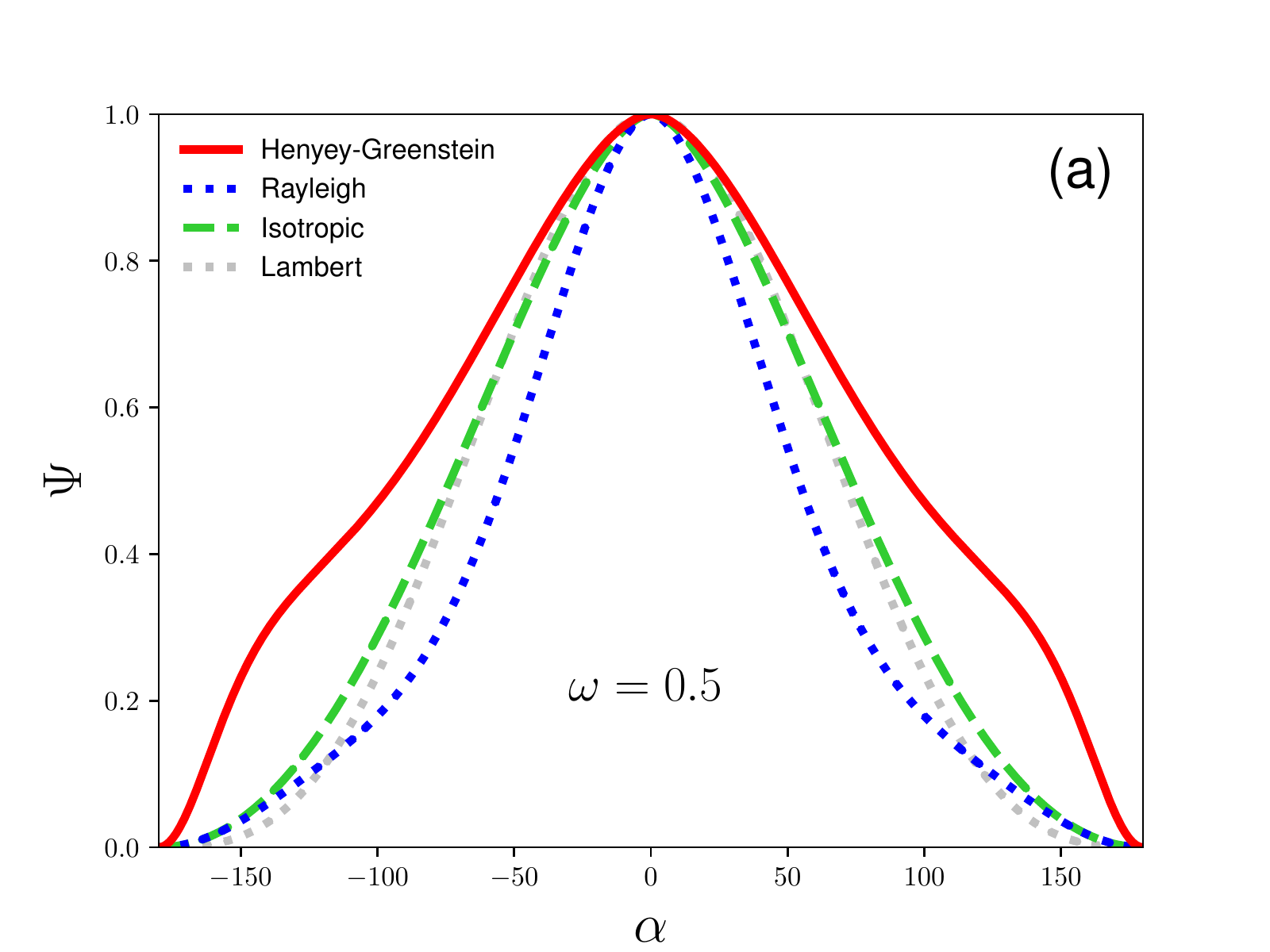}
\includegraphics[width=\columnwidth]{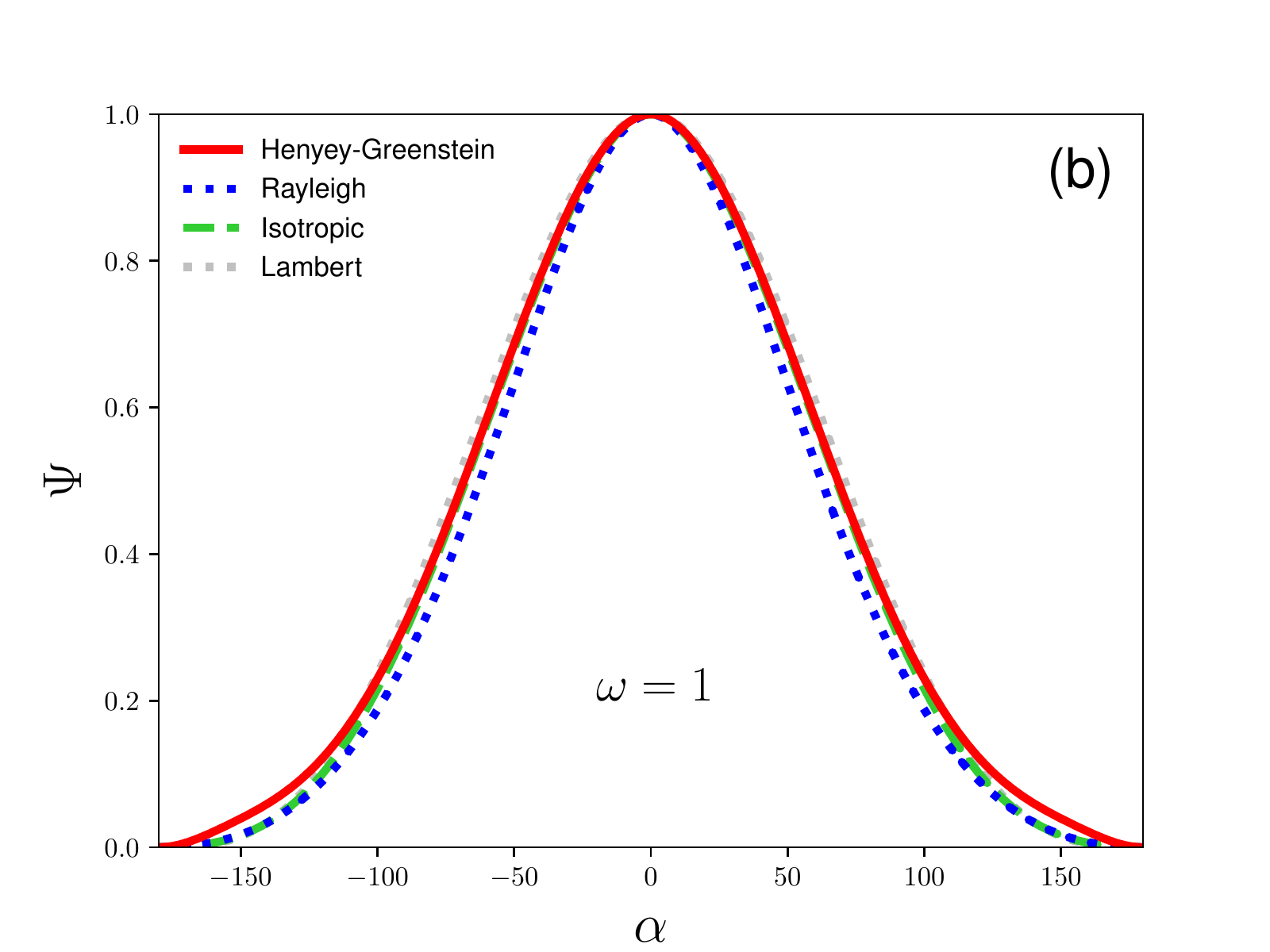}
\end{center}
\vspace{-0.1in}
\caption{Integral phase function for the isotropic, Lambertian, Rayleigh and Henyey-Greenstein reflection laws for (a) $\omega=0.5$ and (b) $\omega=1$.  As the strength of scattering increases, the integral phase function becomes approximately Lambertian.  For illustration, the scattering asymmetry factor of the Henyey-Greenstein reflection law is chosen to be $g=0.508$ such that the phase integral (single scattering only) has a value of 3.5, which matches the ensemble averaged value of a population of hot Jupiters \cite{sc15}.}
\label{fig:psi}
\end{figure*}

\begin{figure*}
\begin{center}
\vspace{-0.2in}
\includegraphics[width=\columnwidth]{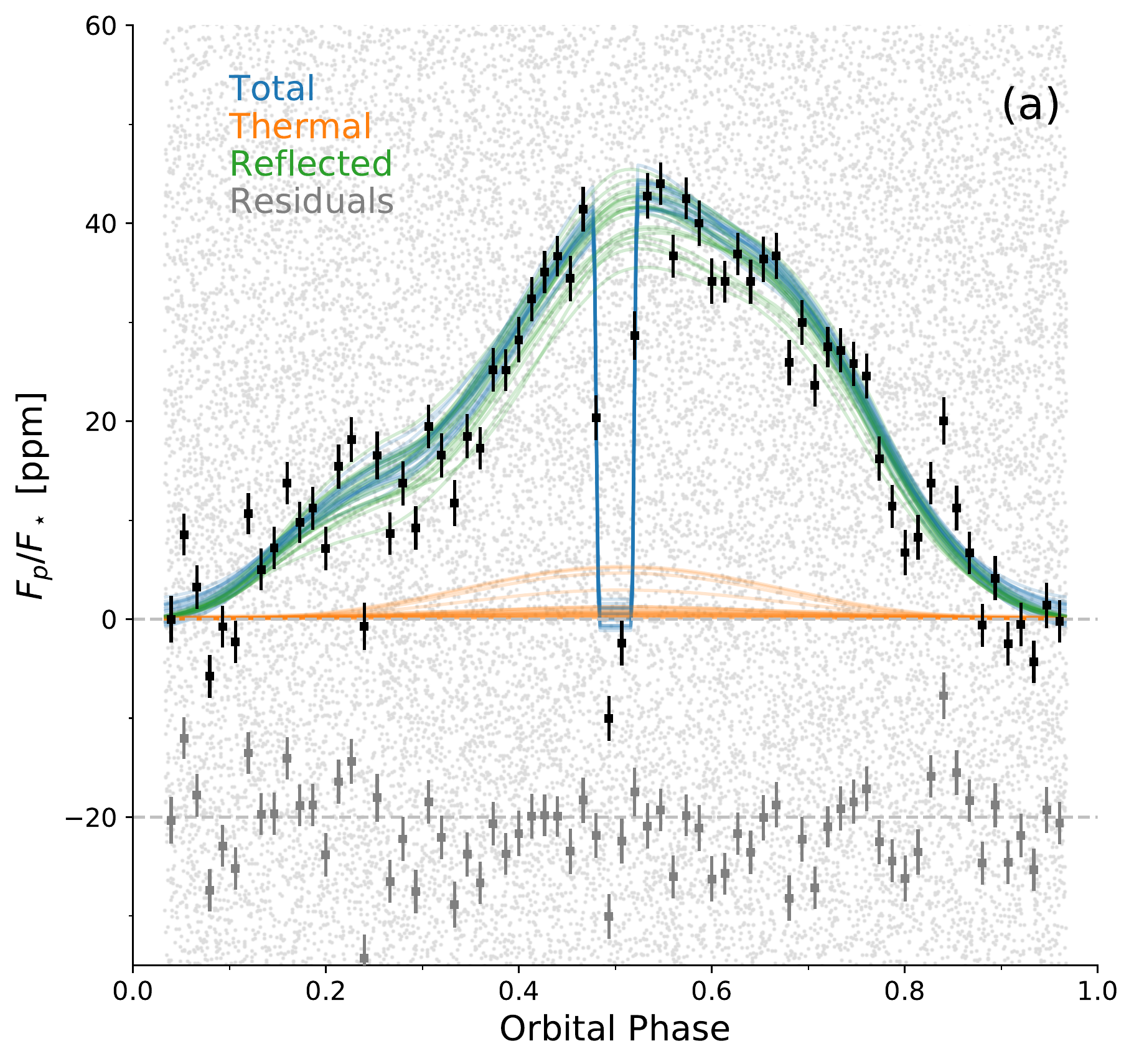}
\includegraphics[width=\columnwidth]{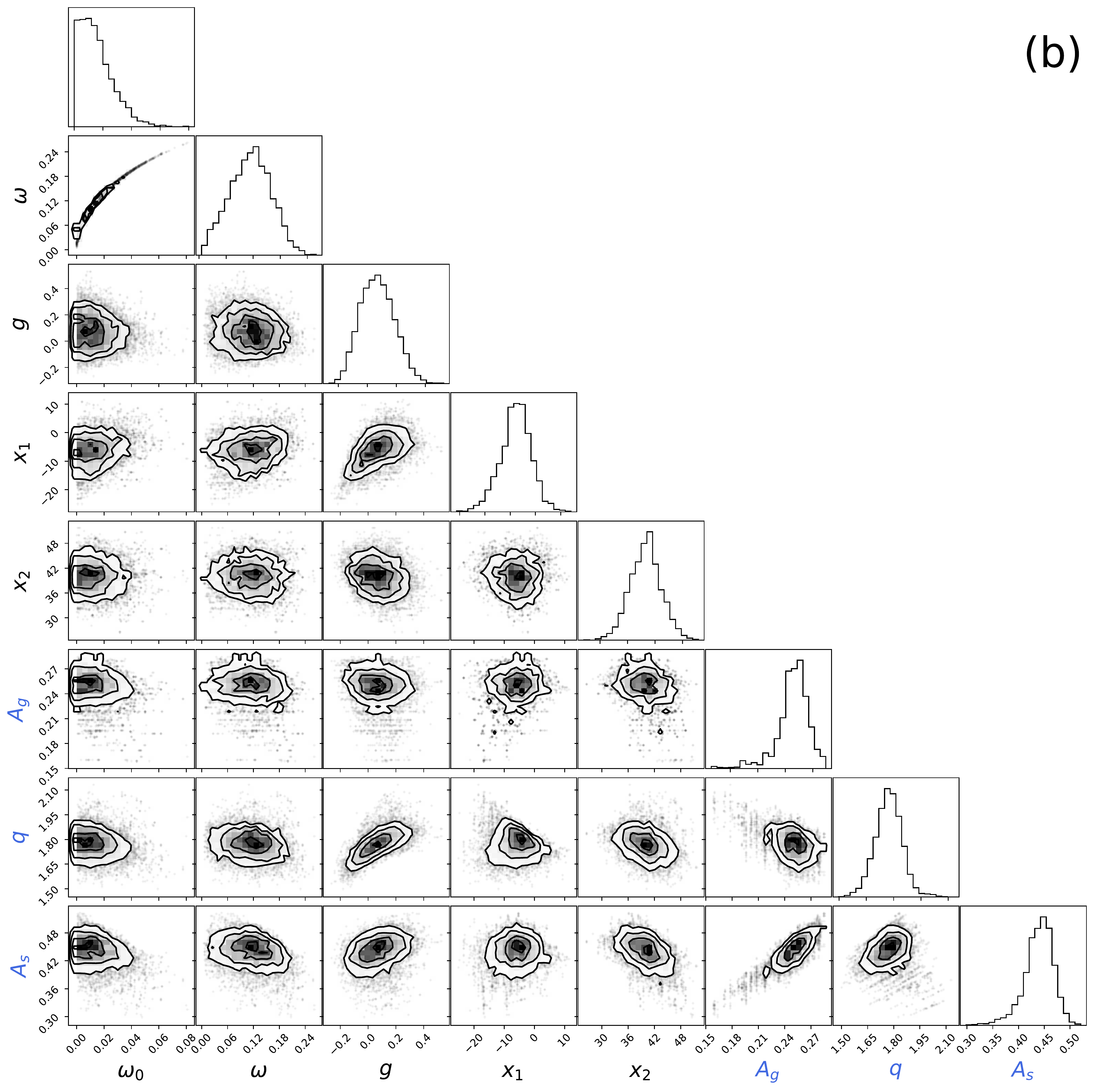}
\end{center}
\vspace{-0.2in}
\caption{(a) We simultaneously fit (blue curve) for the reflected light (green curve), thermal emission (orange curve), secondary eclipse and a stellar/Gaussian process model for the Kepler phase curve of the hot Jupiter Kepler-7b, assuming an atmosphere with inhomogeneous cloud/haze cover. The observations are shown phase-folded and binned only for clarity (black squares).  The fit is performed in the time domain for all 60,000 long-cadence fluxes without binning or phase folding. (b) Posterior distributions of physical parameters (black labels): the baseline single-scattering albedo $\omega_0$, the total single-scattering albedo $\omega$, the scattering asymmetry factor $g$, the local longitudes $x_1$ and $x_2$ between which the single-scattering albedo is $\omega_0$.  Also shown are the posterior distributions for the derived quantities (blue labels), which can be computed from the physical parameters: the geometric albedo $A_g$, the phase integral $q$, and the spherical albedo $A_s$ in the Kepler bandpass.}
\vspace{-0.1in}
\label{fig:kep7b}
\end{figure*}

There are implications of the closed-form solutions of $A_g$ and $\Psi$ for fitting and interpreting the measured reflected light phase curves of exoplanets.  Let the fluxes of the exoplanet and star observed at Earth be $F_p$ and $F_\star$, respectively.  The reflected light phase curve is \cite{seager}
\begin{equation}
\frac{F_p}{F_\star} = \left( \frac{R}{a} \right)^2 A_g \Psi,
\label{eq:phase_curve}
\end{equation}
where $R$ is the radius of the exoplanet and $a$ is its orbital semi-major axis.  Figure \ref{fig:psi} shows several examples of $\Psi$.  At low and intermediate values of the single-scattering albedo $\omega$, the shapes of reflected light phase curves encode information on the reflection laws.  As $\omega$ approaches unity, the phase curve shape follows that of a Lambertian sphere \cite{madhu12}.  A severe test of the theory is to measure reflected light phase curves of cloudfree exoplanetary atmospheres, where Rayleigh scattering of light is mediated by atoms and molecules and the phase curve is predicted to have a distinct shape if $\omega<1$ (Figure \ref{fig:psi}).

We use the closed-form solutions to fit the phase curve of the hot Jupiter Kepler-7b measured by the Kepler space telescope.  Since the reflected light component of the phase curve has a peak offset \cite{demory13}, a model of an atmosphere with inhomogeneous cloud/haze cover \cite{hu15,sh15} is needed to fit the data (see Methods).  For tidally-locked hot Jupiters, an inhomogeneous atmosphere is a natural consequence of aerosols having finite condensation temperatures \cite{oreshenko16}. It is assumed that the atmosphere has a single-scattering albedo $\omega_0$ between two local longitudes $x_1$ and $x_2$, while its other regions are enhanced by $\omega^\prime$ such that the total single-scattering albedo is $\omega = \omega_0 + \omega^\prime$ (Figure \ref{fig:inhomogeneous}). The asymmetry of scattering is globally quantified by the scattering asymmetry factor $g$ (see Methods).

Standard phase curve analyses perform detrending of the raw telescope measurements and fitting to a phase curve model as separate steps \cite{esteves15,demory13}.  Often after detrending, the photometry is binned in time to reduce data volume and increase the signal-to-noise of the weak phase curve signal.  A key improvement from previous analyses is the ability to jointly and self-consistently fit the shape of the phase curve and secondary eclipse depth in terms of physical parameters ($\omega_0$, $\omega$, $g$, $x_1$ and $x_2$).  A major implication is the ability to simultaneously detrend the phase curve in the time domain \textit{and} fit for physical parameters at the \textit{native} temporal resolution of the photometry---without the need to bin in time (see Methods).  Performing both steps simultaneously and without binning \cite{kipping10} ensures that uncertainties in the detrending process are accurately propagated into the physical parameters.  

Figure \ref{fig:kep7b} shows the model fit to the Kepler-7b phase curve, which indicate that the atmosphere has a dark region with a single-scattering albedo of $\omega_0={0.0136}^{+0.0132}_{-0.0093}$ between the local longitudes of $x_1={-6.0}^{+4.4}_{-5.0}$ and $x_2={40.0}^{+3.3}_{-3.4}$ degrees.  The reflective regions of the atmosphere have a single-scattering albedo of $\omega={0.115}^{+0.044}_{-0.049}$ and a scattering asymmetry factor of $g={0.07}^{+0.12}_{-0.11}$, the latter of which is consistent with isotropic scattering of starlight by aerosols with sizes smaller than the wavelengths probed by the Kepler bandpass (0.42--0.90 $\mu$m). The bandpass-integrated geometric albedo is $A_g={0.251}^{+0.012}_{-0.016}$, which is consistent with a previous study \cite{esteves15} but revises the value of $A_g = 0.35 \pm 0.02$ reported in the discovery paper \cite{demory13}.  Using the retrieved values of $\omega_0$, $\omega$, $g$, $x_1$ and $x_2$, the Kepler bandpass-integrated phase integral and spherical albedo are inferred to be $q={1.774}^{+0.068}_{-0.073}$ and $A_s=0.442^{+0.023}_{-0.028}$, respectively. The transition between the dark and reflective regions of the atmosphere occurs at $\sim 1600$ K, which we interpret as the condensation temperature of the aerosols.  At solar metallicity, a variety of aerosol compositions are plausible \cite{burrows06}.

The closed-form, ab initio solutions presented here are of intermediate complexity and fill an important gap between ad hoc models and complex numerical calculations. They enable novel data analyses to be performed; for example, to retrieve the globally averaged properties of aerosols in the atmosphere of Jupiter from phase curves measured by the Cassini spacecraft \cite{hl21}.  In the upcoming era of the James Webb Space Telescope, phase curves of exoplanets from 0.6 to 24 $\mu$m will be procured.  The closed-form solutions presented here enable physical parameters to be extracted from multi-wavelength phase curves within a Bayesian framework, which will motivate further study using complex numerical simulations. 

\vspace{0.2in}

\noindent
{\scriptsize \textbf{Data Availability:} The data that support the plots within this paper and other findings of this study are available from the corresponding author upon reasonable request. 

\vspace{0.2in}

\noindent
\textbf{Code Availability:} The code used to compute the models and perform Bayesian inference is available at \texttt{https://github.com/bmorris3/kelp}.

\vspace{0.2in}

\noindent
Correspondence and request for materials should be made to K.H.  We acknowledge partial financial support from the Center for Space and Habitability (K.H., B.M.M. and D.K.), the PlanetS National Center of Competence in Research (B.M.M.) and an European Research Council (ERC) Consolidator Grant awarded to K.H. (project EXOKLEIN; number 771620). K.H. acknowledges a honorary professorship from the Department of Physics of the University of Warwick and an imminent chair professorship of theoretical astrophysics from the Ludwig Maximilian University.}

\vspace{0.2in}

\noindent
{\scriptsize K.H. formulated the problem, combined insights from the historical literature, derived the equations, made all of the figures except Figure 4 and led the writing of the manuscript.  B.M.M. designed and authored open-source software that implemented the derived equations, performed the analysis of Kepler-7b data, made Figure 4 and co-wrote the manuscript.  D.K. participated in decisive discussions of the problem with K.H. and read the manuscript.}

\setcounter{figure}{0}
\renewcommand{\figurename}{Extended Data Figure}
\section*{Methods}

\subsection*{Spherical trigonometry}

The six angles of the local and observer-centric coordinate systems (Figure \ref{fig:geometry}) are related by spherical trigonometry \cite{sobolev},
\begin{equation}
\begin{split}
\mu &= \cos \Theta ~\cos\Phi , \\
\mu_\star &= \cos\Theta ~\cos\left( \alpha - \Phi \right), \\
\cos\alpha &= \mu \mu_\star - \sqrt{\left(1-\mu^2\right) \left(1-\mu^2_\star \right)} ~\cos\phi.
\end{split}
\end{equation}
At zero phase angle ($\alpha=0$), one obtains $\mu=\mu_\star$ and $\phi=\pi$ \cite{sobolev}.

\subsection*{Radiative transfer equation} 

Let the scattering phase function be represented by $P$, which describes the mathematical relationship between the incident and emergent angles of radiation.  It is chosen to have no physical units; an alternative choice is for it to have units of per unit solid angle.  The fraction of light reflected during a single scattering event is given by the single-scattering albedo $\omega$.  With these definitions, the radiative transfer equation reads \cite{sobolev,chandra,pierrehumbert,mihalas,toon89,heng,heng18}
\begin{equation}
\mu \frac{\partial I}{\partial \tau} = I - \frac{\omega}{4\pi} \int^{4\pi}_0 I P ~d\Omega^\prime - \frac{\omega I_\star}{4} P_\star e^{-\tau/\mu_\star}.
\label{eq:rt_equation}
\end{equation}
The optical depth $\tau$ is the generalisation of length in radiative transfer and takes into account not just the spatial extent of the atmosphere, but how dense it is and the ability of its constituent atoms, molecules, ions and aerosols to absorb and scatter radiation.  Since we are interested in calculating the intensity of reflected light, equation (\ref{eq:rt_equation}) ignores the contribution of thermal (blackbody) emission.  The term involving the integral accounts for the multiple scattering of radiation, while the last term describes the collimated beam of incident starlight.  The integral is generally difficult to evaluate, because the integration is performed over all incident angles of radiation (with $d\Omega^\prime$ denoting the incident solid angle).  

The scattering phase function is normalised such that \cite{pierrehumbert,heng}
\begin{equation}
\frac{1}{4\pi}\int^{4\pi}_0 P ~d\Omega = \frac{1}{4\pi}\int^{4\pi}_0 P ~d\Omega^\prime = 1,
\end{equation}
with $d\Omega$ denoting the emergent solid angle.  It depends on both the incident and emergent angles: $P(\mu^\prime, \mu, \phi)$.  Azimuthal symmetry is assumed.  For the stellar beam, we have $P_\star = P(-\mu_\star, \mu, \phi)$; the minus sign accounts for the convention chosen that radiation travelling into the atmospheric column, towards the center of the exoplanet, corresponds to $\mu^\prime<0$.  At zero phase angle, the scattering phase function becomes $P_0 = P(-\mu_\star, \mu_\star, \pi)$.  The scattering asymmetry factor is the first moment of the scattering phase function \cite{pierrehumbert,heng},
\begin{equation}
g = \frac{1}{4\pi} \int^{2\pi}_0 \int^\pi_0 P ~\cos\beta ~\sin\beta ~d\beta ~d\phi.
\end{equation}
When $P=1$ (isotropic scattering), one naturally obtains $g=0$.

\subsection*{Solutions of radiative transfer equation} 

If one ignores the integral in equation (\ref{eq:rt_equation}), then only single scattering is considered.  For a semi-infinite atmosphere, the intensity of reflected light is \cite{hapke81}
\begin{equation}
I_{0,{\rm S}} = \frac{\omega I_\star}{4} \frac{\mu_\star}{\mu_\star + \mu} P_\star.
\end{equation}
In the limit of isotropic single scattering ($P_\star=1$), one obtains the Lommel-Seeliger law of reflection \cite{lommel,seeliger}.  For isotropic multiple scattering, Chandrasekhar's exact solution ignoring the stellar beam is \cite{horak50,chandra}
\begin{equation}
I_{0,{\rm M}} = \frac{\omega I_\star}{4} \frac{\mu_\star}{\mu_\star + \mu} H H_\star,
\end{equation}
where $H_\star = H(\mu_\star)$ and $H(\mu)$ is the Chandrasekhar H-function \cite{chandra},
\begin{equation}
H = 1 + \frac{1}{2} \omega \mu H \int^1_0 \frac{H_\star}{\mu_\star + \mu} ~d\mu_\star.
\end{equation}
An identical solution for $I_{0,{\rm M}}$ may be found by evaluating the integral in equation (\ref{eq:rt_equation}) using the two-stream solutions, but with the H-function taking on an approximate form (Hapke's solution) \cite{hapke81,davidovic08},
\begin{equation}
H = \frac{1 + 2\mu}{1 + 2\gamma \mu},
\end{equation}
where we have $\gamma = \sqrt{1-\omega}$.  To combine the solutions, one needs to account for isotropic multiple scattering of the stellar beam, which yields \cite{hapke81}
\begin{equation}
I_0 = \frac{\omega I_\star}{4} \frac{\mu_\star}{\mu_\star + \mu} \left( P_\star - 1 + H H_\star \right).
\end{equation}
The reflection coefficient is obtained simply using $\rho = I_0/I_\star \mu_\star$.

\subsection*{Formulae of albedos} 

If only single scattering is considered, it is straightforward to derive the geometric albedo,
\begin{equation}
A_{g,{\rm S}} = \frac{\omega P_0}{8}.
\end{equation}
The derivation may be performed in either the local or observer-centric coordinate systems.  To evaluate the geometric albedo for isotropic multiple scattering, one uses Hapke's linear approximation of the Chandrasekhar H-function \cite{hapke81},
\begin{equation}
H = \left( 1 + \epsilon \right) \left( 1 - \frac{\epsilon}{2} + \epsilon \mu \right).
\label{eq:h_function_linear}
\end{equation}
The geometric albedo associated with isotropic multiple scattering only is
\begin{equation}
\begin{split}
A_{g,{\rm M}} &= \frac{\omega}{4} \int^1_0 \mu H^2 ~d\mu - \frac{\omega}{8} \\
&= \frac{\epsilon}{2} + \frac{\epsilon^2}{6} + \frac{\epsilon^3}{24} - \frac{\omega}{8}.
\end{split}
\end{equation}
The total geometric albedo is $A_g = A_{g,{\rm S}}  + A_{g,{\rm M}}$.

There is an alternative way of deriving the spherical albedo without having to evaluate the phase integral $q$ in equation (\ref{eq:phase_integral}).  In Figure \ref{fig:geometry}, the flux reflected by the infinitesimal area element with projected solid angle $\mu ~d\Omega$ is $\int I_0 \mu ~d\Omega = \int \rho I_\star \mu_\star \mu ~d\Omega$.  In the local coordinate system, one has $d\Omega = d\mu ~d\phi$.  The incident flux of starlight is $\pi \mu_\star I_\star$.  Therefore, the plane albedo associated with each atmospheric column is \cite{sobolev}
\begin{equation}
A = \frac{1}{\pi} \int^{2\pi}_0 \int^1_0 \rho \mu ~d\mu ~d\phi,
\end{equation}
where $0 \le \mu \le 1$ represents only the emergent (and not downwelling) radiation from the exoplanet.  The term ``plane albedo" refers to each local patch of atmosphere being approximated as a plane-parallel layer \cite{sobolev}.  The fraction of incident starlight reflected by the entire exoplanet needs to consider all of the atmospheric columns.  The reflected flux is $2 \pi \int \mu_\star \pi I_\star A ~d\mu_\star$, while the incident flux is $\pi^2 I_\star$.  The ratio of these two quantities yields the spherical albedo \cite{sobolev},
\begin{equation}
A_s = 2 \int^1_0 A \mu_\star ~d\mu_\star.
\end{equation}
The isotropic multiple scattering component of the spherical albedo is \cite{hapke81}
\begin{equation}
A_{s,{\rm M}} = \frac{5 \epsilon}{6} + \frac{\epsilon^2}{6} - \frac{2}{3} \omega \left( 1 - \ln{2} \right),
\end{equation}
where Hapke's linear approximation of the Chandrasekhar H-function has again been invoked.  Finally, the spherical albedo including both single and isotropic multiple scattering is
\begin{equation}
A_s = \frac{q_{\rm S} \omega P_0}{8} + A_{s,{\rm M}},
\end{equation}
where the single scattering phase integral $q_{\rm S}$ is obtained from
\begin{equation}
q_{\rm S} = \frac{2}{P_0} \int^{\pi}_0 P_\star \Psi_{\rm S} ~\sin\alpha ~d\alpha.
\end{equation}
For isotropic single scattering ($P_\star = P_0 = 1$), we verified numerically that $q_{\rm S} = 16(1-\ln2)/3 \approx 1.63655$ \cite{russell1916,hh89}.  For Rayleigh scattering, the single-scattering component of the spherical albedo may be derived exactly,
\begin{equation}
\begin{split}
A_{s,{\rm S}} =& \frac{3\omega}{8} \int^1_0 \int^1_0 \frac{\mu_\star\mu}{\mu_\star + \mu} \left( 3 + 3 \mu_\star^2 \mu^2 \right. \\
&\left. - \mu_\star^2 - \mu^2 \right) ~d\mu ~d\mu_\star = \frac{\omega}{35}\left( 26 -27\ln{2} \right),
\end{split}
\end{equation}
which has a value of about $0.208144 \omega$.  We verified numerically that $q_{\rm S} \omega P_0/8\omega \approx 0.208144$.  

In the limit of conservative ($\omega=1$) Rayleigh single scattering and isotropic multiple scattering, we obtain
\begin{equation}
\frac{A_g}{A_s} = \frac{1295}{1808 - 176 \ln{2}} \approx 0.768,
\end{equation}
which is not inconsistent with the estimate of 0.751 for conservative Rayleigh scattering \cite{dy74} that is the basis for the commonly quoted factor of 3/4 \cite{marley99,madhu12}.  The slight discrepancy of about 2.3\% probably stems from the assumption of isotropic multiple scattering.

\subsection*{Dimensionless Sobolev fluxes}

\begin{figure*}
\begin{center}
\vspace{-0.35in}
\includegraphics[width=0.95\columnwidth]{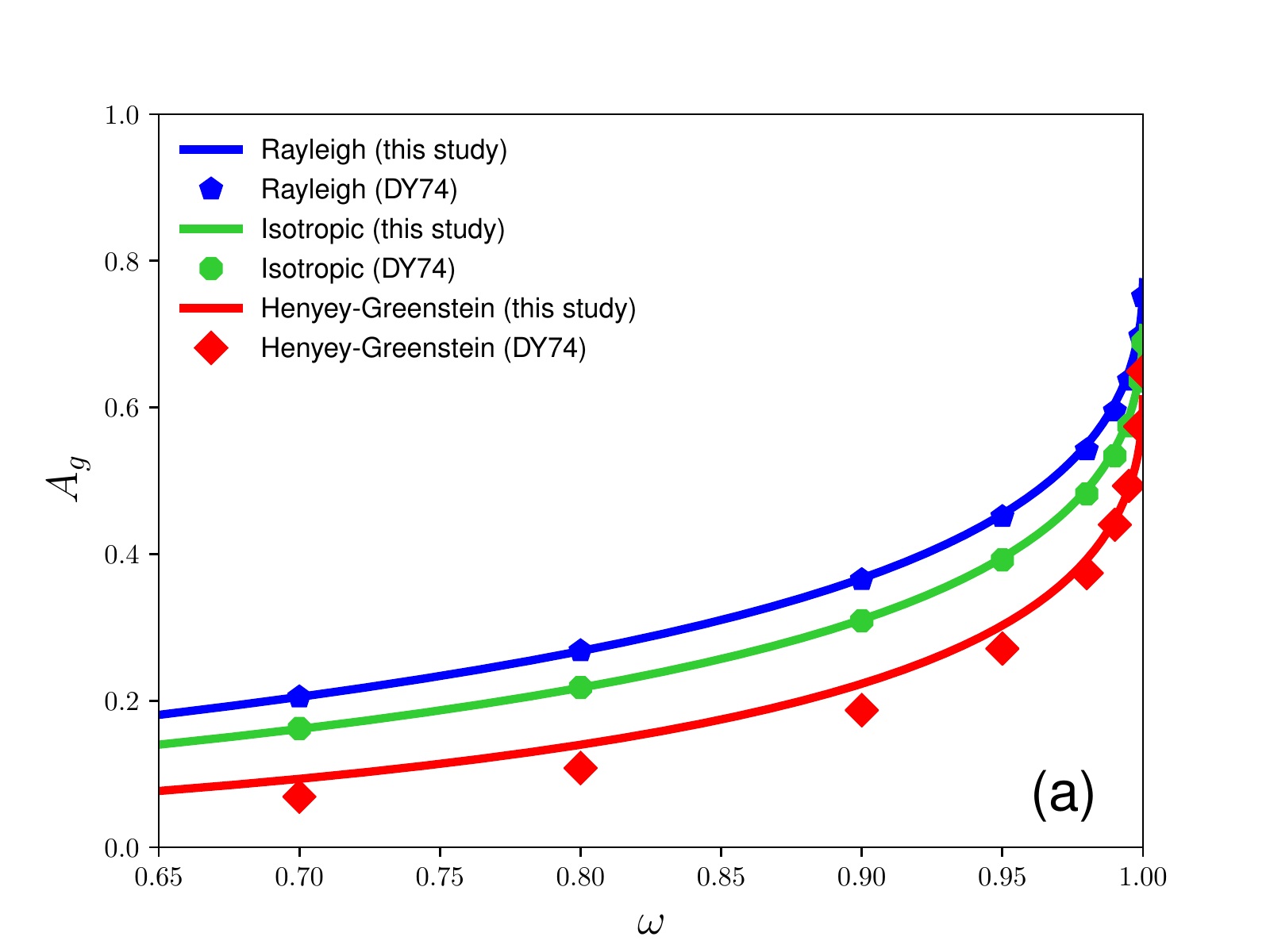}
\includegraphics[width=0.95\columnwidth]{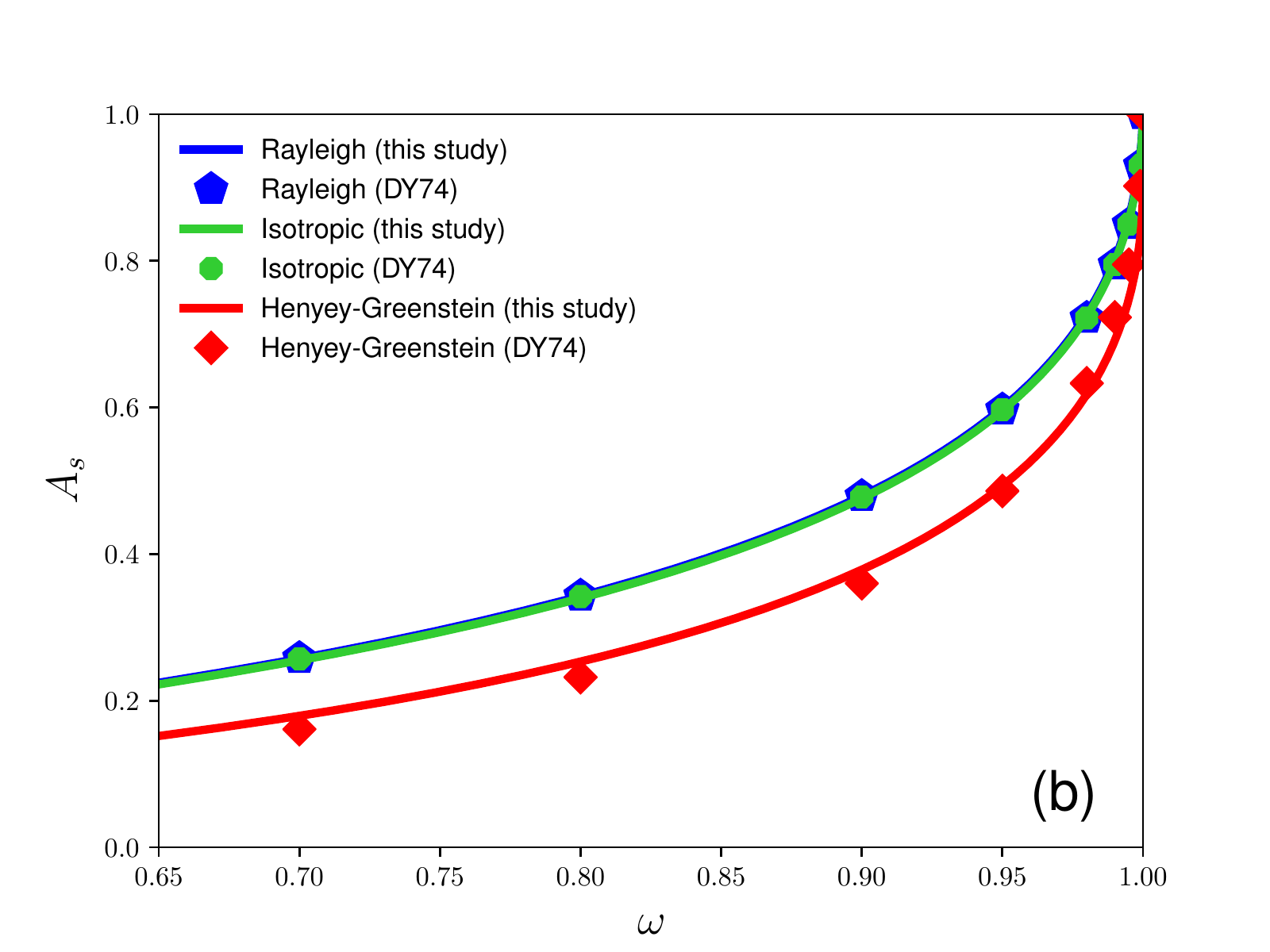}
\includegraphics[width=0.9\columnwidth]{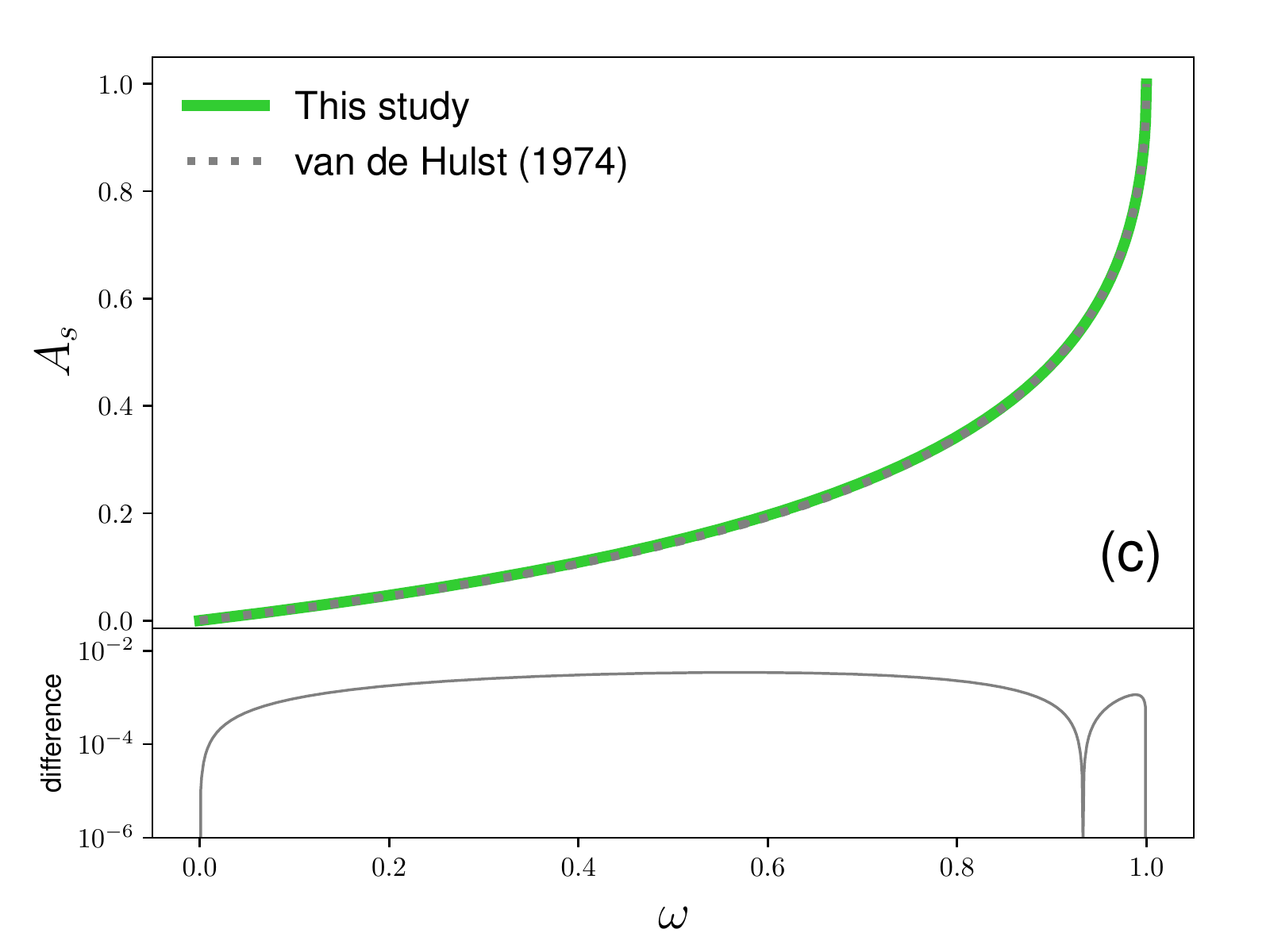}
\includegraphics[width=0.9\columnwidth]{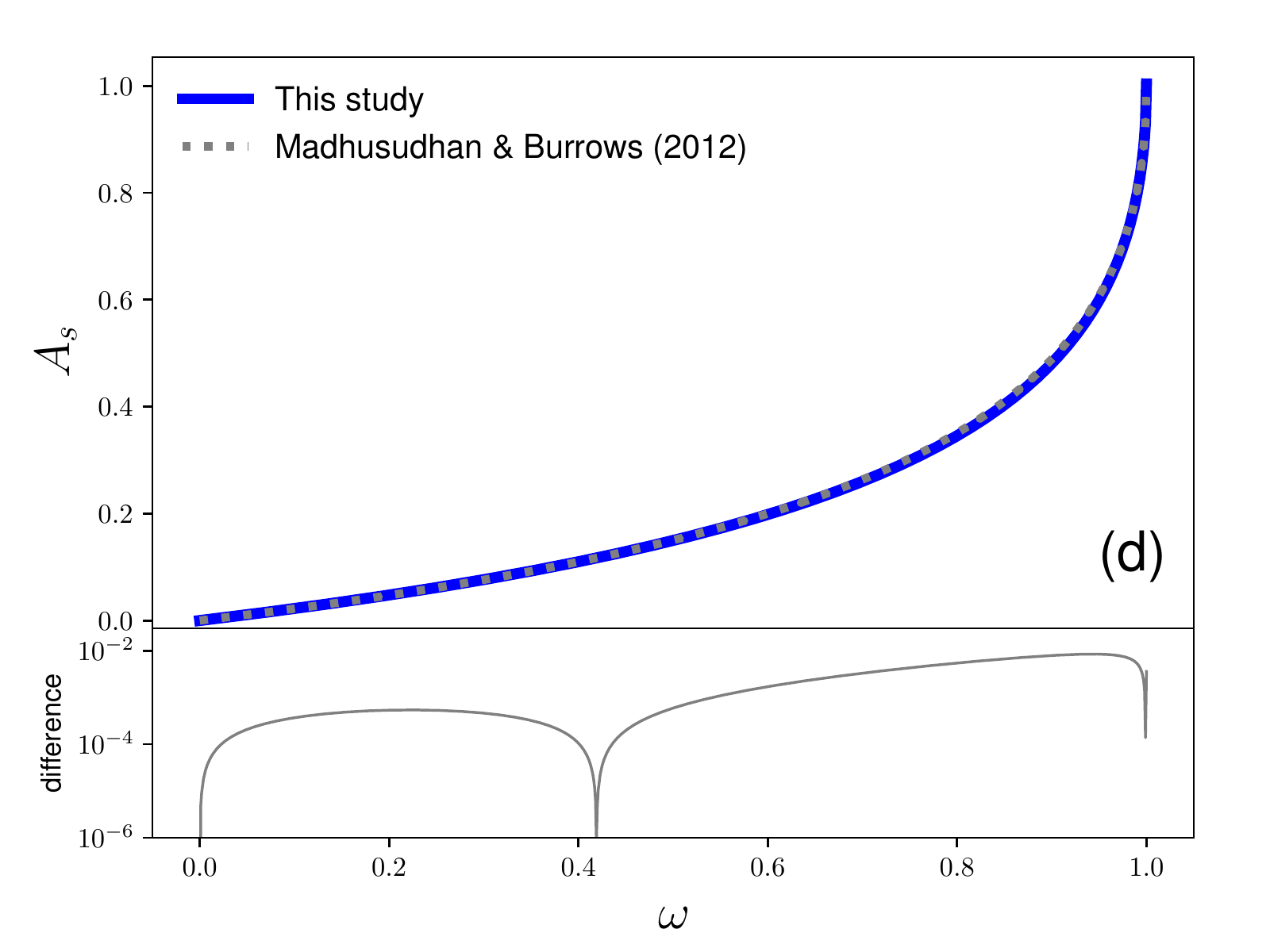}
\end{center}
\vspace{-0.1in}
\caption{\scriptsize Validation of calculations against the classic work of Dlugach \& Yanovitskij (labelled DY74) \cite{dy74} for the (a) geometric albedo and (b) spherical albedo for various reflection laws.  For the Henyey-Greenstein reflection law, the scattering asymmetry factor assumed is $g=0.5$.  For isotropic scattering, (c) shows the comparison of the spherical albedo to the classic work of van de Hulst \cite{van74}. For non-conservative Rayleigh scattering, (d) shows the comparison of the spherical albedo to the work of Madhusudhan \& Burrows \cite{madhu12}.}
\label{fig:benchmark}
\end{figure*}

The integral phase function is constructed from the fluxes, which are rendered dimensionless by dividing them by $I_\star$ \cite{sobolev}.  Here, we write down the key steps involved in deriving these fluxes.  As stated in equation (\ref{eq:flux_sobolev}), the fluxes are constructed from the reflection coefficient $\rho$.  Using Hapke's linear approximation of the Chandrasekhar H-function \cite{hapke81}, the reflection coefficient may be written as
\begin{equation}
\rho = \frac{\omega}{4} \left( \frac{\rho_{\rm S}}{\mu_\star + \mu} + \rho_{\rm L} + \rho_{\rm C} \frac{\mu_\star \mu}{\mu_\star + \mu}\right),
\end{equation}
where the various coefficients have already been defined in equation (\ref{eq:rho_coefficients}).  The dimensionless flux of Sobolev may be written as three terms,
\begin{equation}
F = F_{\rm S} + F_{\rm L} + F_{\rm C}.
\end{equation}
For a homogeneous atmosphere, the term that exhibits single scattering behaviour is
\begin{equation}
\begin{split}
F_{\rm S} =& \frac{\omega \rho_{\rm S}}{4}  \int^{\pi/2}_0 \cos^2\Theta ~d\Theta \\
&\times \int^{\pi/2}_{\alpha-\pi/2} \frac{\cos\Phi ~\cos\left(\alpha-\Phi\right)}{\cos\Phi + \cos\left(\alpha-\Phi\right)} ~d\Phi.
\end{split}
\end{equation}
The Lambertian-sphere-like term is
\begin{equation}
\begin{split}
F_{\rm L} =& \frac{\omega \rho_{\rm L}}{4}  \int^{\pi/2}_0 \cos^3\Theta ~d\Theta \\
&\times \int^{\pi/2}_{\alpha-\pi/2} \cos\Phi ~\cos\left(\alpha-\Phi\right) ~d\Phi.
\end{split}
\end{equation}
The correction term is
\begin{equation}
\begin{split}
F_{\rm C} =& \frac{\omega \rho_{\rm C}}{4}  \int^{\pi/2}_0 \cos^4\Theta ~d\Theta \\
&\times \int^{\pi/2}_{\alpha-\pi/2} \frac{\cos^2\Phi ~\cos^2\left(\alpha-\Phi\right)}{\cos\Phi + \cos\left(\alpha-\Phi\right)} ~d\Phi.
\end{split}
\end{equation}
Two key properties of all three terms are worth belabouring: the integrals over $\Theta$ and $\Phi$ may be evaluated independently if $P=P(\beta)$, and closed-form solutions exist for them.  By defining $F_0=F(\alpha=0)$, the integral phase function of a homogeneous atmosphere may be obtained using $\Psi=F/F_0$.

\subsection*{Henyey-Greenstein reflection law}

The widely used Henyey-Greenstein scattering phase function is \cite{hg}
\begin{equation}
P_\star = \frac{1 - g^2}{\left( 1 + g^2 - 2 g \cos \beta \right)^{3/2}},
\end{equation}
where the scattering angle $\beta$ is \cite{pierrehumbert,heng,heng18}
\begin{equation}
\cos\beta = -\mu \mu_\star + \sqrt{\left(1-\mu^2\right) \left(1-\mu^2_\star \right)} ~\cos\phi.
\end{equation}
The scattering asymmetry factor $g$ quantifies the asymmetry of scattering: $g=-1$, 0 and 1 correspond to purely reverse, isotropic and purely forward scattering, respectively.  The scattering phase function at zero phase angle ($\cos\beta = -\cos\alpha =-1$) is
\begin{equation}
P_0 = \frac{1 - g}{\left( 1 + g \right)^2}.
\end{equation}

\subsection*{Validation}

\begin{figure}
\begin{center}
\vspace{-0.2in}
\includegraphics[width=\columnwidth]{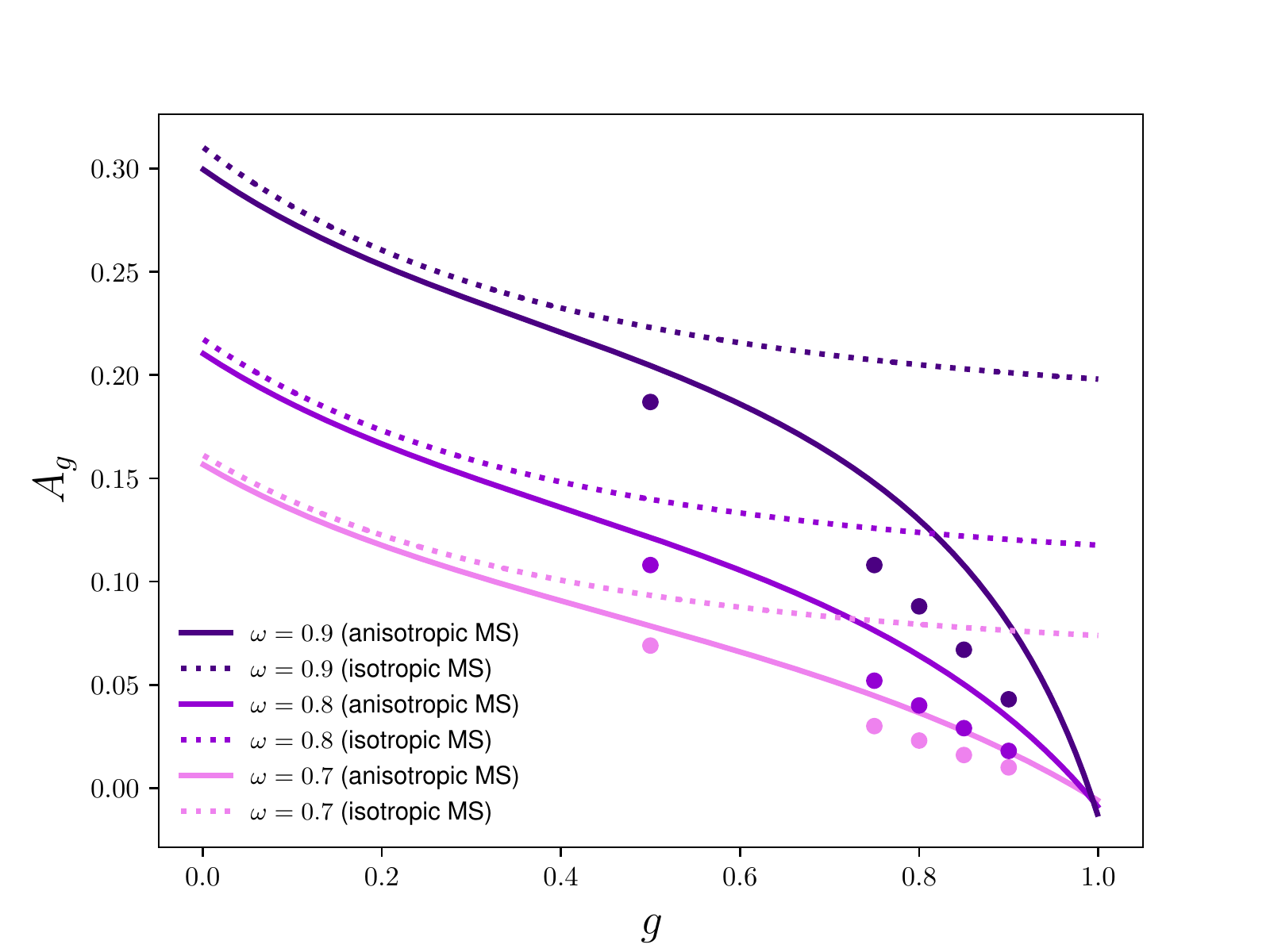}
\end{center}
\vspace{-0.1in}
\caption{\scriptsize Calculations of the geometric albedo comparing anisotropic versus isotropic multiple scattering (MS), which follow Hapke's approach of utilising the two-stream fluxes. Overlaid as circles are the calculations of Dlugach \& Yanovitskij \cite{dy74}. The discrepancies at low values of $g$ originate from the use of Hapke's linear approximation to the Chandrasekhar H-function for isotropic multiple scattering \cite{hapke81}.}
\label{fig:benchmark2}
\end{figure}

To validate our calculations of $A_g$ and $A_s$, we compare them against the numerical calculations of Dlugach \& Yanovitskij \cite{dy74}, who tabulated the geometric and spherical albedos from $\omega=0.7$ to 1  (Extended Data Figure \ref{fig:benchmark}).  For isotropic and Rayleigh scattering, the discrepancies associated with $A_g$ and $A_s$ are typically $\sim 0.1$--1\%.  For the Henyey-Greenstein law of reflection, the discrepancies associated with $A_g$ and $A_s$ are $\sim 10\%$ for $g=0.5$.  The larger discrepancies indicate that the assumption of isotropic multiple scattering starts to break down when scattering becomes strongly asymmetric.  The error becomes large ($\sim 100\%$) for $g>0.5$, implying that the isotropic multiple scattering assumption should not be used (Extended Data Figure \ref{fig:benchmark2}).  For isotropic scattering and non-conservative ($\omega \ne 1$) Rayleigh scattering, we perform further comparisons of the spherical albedo to the empirical fitting functions of van de Hulst \cite{van74} and Madhusudhan \& Burrows \cite{madhu12}, respectively.  The differences between these fitting functions and our calculations are negligible ($<1\%$), as shown in Extended Data Figure \ref{fig:benchmark}.

\begin{figure}
\begin{center}
\vspace{-0.2in}
\includegraphics[width=\columnwidth]{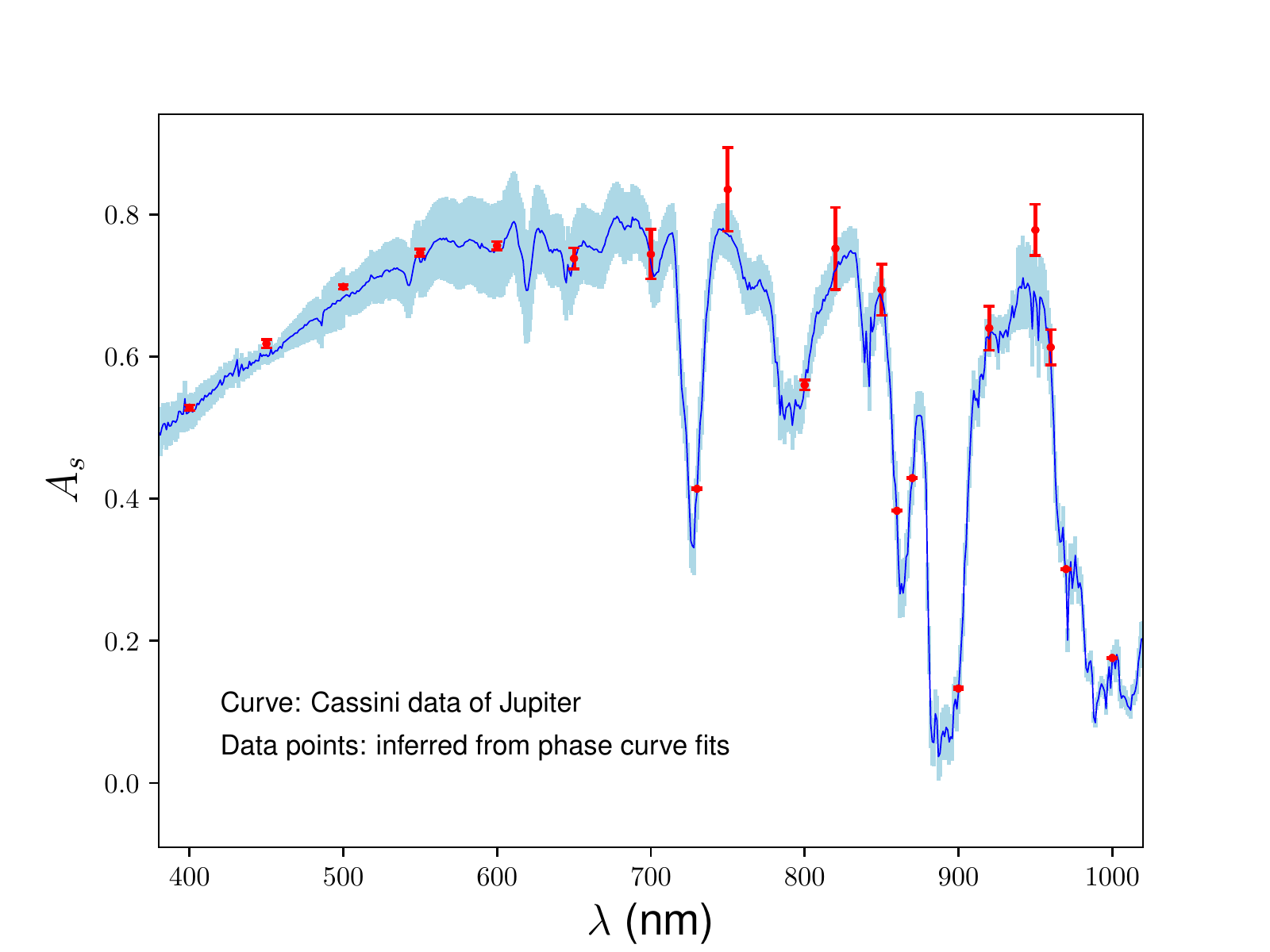}
\end{center}
\vspace{-0.1in}
\caption{\scriptsize Comparing measurements of the spherical albedo of Jupiter (curve with 1$\sigma$ uncertainties) with the values inferred from phase curve fitting (circles with 1$\sigma$ uncertainties). All of the data used were measured by the Cassini spacecraft.}
\label{fig:jupiter}
\end{figure}

A different form of validation concerns the inference of the spherical albedo from phase curve fitting. The globally averaged values of the geometric albedo, phase integral and spherical albedo have been measured for Jupiter of the Solar System by the Cassini spacecraft \cite{li18}.  A previous study fitted $A_g \Psi$ with a double Henyey-Greenstein reflection law to Cassini phase curves of Jupiter from 0.4--1.0 $\mu$m \cite{hl21}. From the fitted $\Psi$, we compute $q$ by propagating the uncertainties in the fitting parameters. The spherical albedo is computed using $A_s = q A_g$, where $A_g$ is the measured geometric albedo; the uncertainties on $q$ and $A_g$ are assumed to be uncorrelated.  Extended Data Figure \ref{fig:jupiter} shows that the values of $A_s$ inferred in this way are consistent with those reported previously \cite{li18}.

\subsection*{Inhomogeneous atmosphere}

\begin{figure}
\begin{center}
\includegraphics[width=\columnwidth]{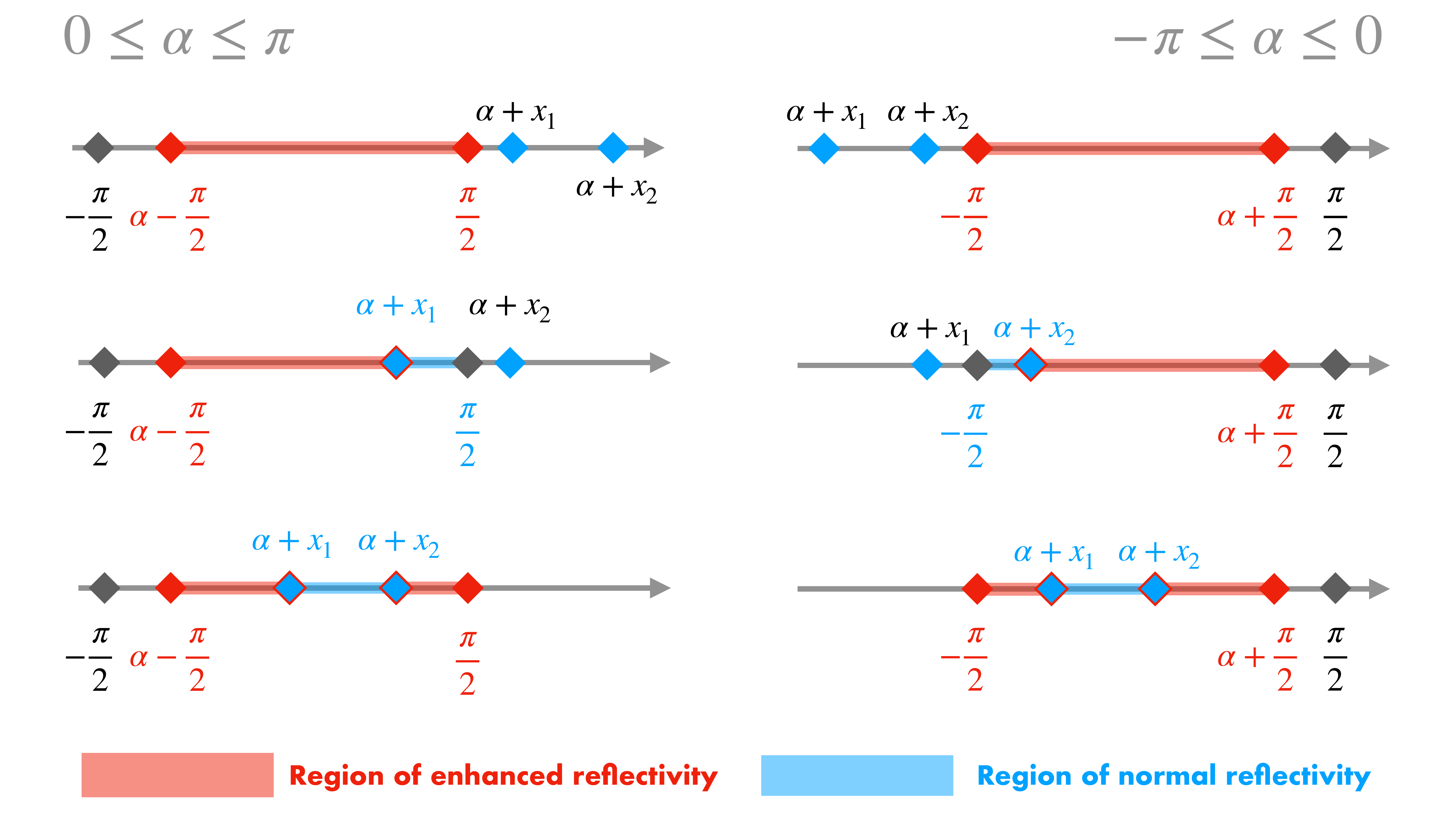}
\end{center}
\caption{\scriptsize Regions of normal and enhanced reflectivity for an inhomogeneous atmosphere in terms of the longitude $\Phi$ within the observer-centric coordinate system.  In the local longitude of the exoplanet (where the substellar point sits at $x=0$), the atmosphere has a baseline single-scattering albedo of $\omega_0$ across $x_1 \le x \le x_2$.  When this region is within view of the observer, it is highlighted with a thick blue line.  Regions of enhanced reflectivity (with a total single-scattering albedo of $\omega=\omega_0+\omega^\prime$) that are within the observer's view are highlighted with thick red lines.}
\label{fig:inhomogeneous}
\end{figure}

Let the local longitude of the exoplanet be $x$ with $x=0$ corresponding to the substellar point.  Since it lies in the same plane as $\alpha$, it is trivially related to the phase angle by $x=\Phi-\alpha$ \cite{hu15}.  It is assumed that the region from $x_1 \le x \le x_2$ has a baseline single-scattering albedo of $\omega_0$, while all of the other regions have an enhanced single-scattering albedo of $\omega_0+\omega^\prime$ \cite{hu15}.  The dimensionless Sobolev fluxes are generalised to
\begin{equation}
\begin{split}
F_{\rm S} &= \frac{\pi}{16} \left( \omega_0 \rho_{\rm S_0} \Psi_{\rm S} + \omega^\prime \rho^\prime_{\rm S} \Psi^\prime_{\rm S} \right), \\
F_{\rm L} &= \frac{\pi}{12} \left( \omega_0 \rho_{\rm L_0} \Psi_{\rm L} + \omega^\prime \rho^\prime_{\rm L} \Psi^\prime_{\rm L} \right), \\
F_{\rm C} &= \frac{3\pi}{64} \left( \omega_0 \rho_{\rm C_0} \Psi_{\rm C} + \omega^\prime \rho^\prime_{\rm C} \Psi^\prime_{\rm C} \right), \\
\end{split}
\end{equation}
where $\rho_{\rm S_0}=\rho_{\rm S}(\omega_0)$, $\rho_{\rm L_0}=\rho_{\rm L}(\omega_0)$, $\rho_{\rm C_0}=\rho_{\rm C}(\omega_0)$, $\rho^\prime_{\rm S} = \rho_{\rm S}\left(\omega^\prime\right)$, $\rho^\prime_{\rm L} = \rho_{\rm L}\left(\omega^\prime\right)$ and $\rho^\prime_{\rm C} = \rho_{\rm C}\left(\omega^\prime\right)$. The coefficients $\rho_{\rm S}$, $\rho_{\rm L}$ and $\rho_{\rm C}$ are given by equation (\ref{eq:rho_coefficients}), while $\Psi_{\rm S}$, $\Psi_{\rm L}$ and $\Psi_{\rm C}$ are given by equation (\ref{eq:Psi}).  The total dimensionless Sobolev flux is $F = F_{\rm S} + F_{\rm L} + F_{\rm C}$, while $F_0 = F(\alpha=0)$ and $A_g = 2 F_0/\pi$.  The integral phase function, $\Psi = F/F_0$, is generally not symmetric about $\alpha=0$ and the generalised expression for the phase integral does not assume this symmetry \cite{hu15},
\begin{equation}
q = \int^{\pi}_{-\pi} \Psi ~\sin{\lvert \alpha \rvert} ~d\alpha.
\end{equation}
To compute $\Psi^\prime_{\rm S}$, $\Psi^\prime_{\rm L}$ and $\Psi^\prime_{\rm C}$, one needs to determine the regions of the inhomogeneous atmosphere that are within view of the observer \cite{hu15} in the observer-centric coordinate system (Extended Data Figure \ref{fig:inhomogeneous}).  If we write $\Psi^\prime_i$ for short (with the index $i={\rm S}, {\rm L}$ or C), then we can express $\Psi^\prime_i$ in terms of the following functions,
\begin{equation}
\begin{split}
I_{\rm S} &= - \frac{1}{2} \sec\left(\frac{\alpha}{2}\right) \left[  \sin\left( \frac{\alpha}{2} - \Phi \right) + \left( \cos\alpha - 1 \right) \Psi^\prime_0 \right], \\
I_{\rm L} &= \frac{1}{\pi} \left[ \Phi \cos\alpha - \frac{1}{2} \sin\left( \alpha - 2\Phi \right) \right], \\
I_{\rm C} &= -\frac{1}{24} \sec\left(\frac{\alpha}{2}\right) \left[ -3\sin\left( \frac{\alpha}{2} - \Phi \right) \right. \\
&\left. + \sin\left( \frac{3\alpha}{2} - 3\Phi \right) + 6\sin\left( \frac{3\alpha}{2} - \Phi \right) \right. \\
&\left. -6\sin\left( \frac{\alpha}{2} + \Phi \right)  + 24\sin^4\left( \frac{\alpha}{2} \right) \Psi^\prime_0 \right], \\
\Psi^\prime_0 &= \tanh^{-1}\left[ \sec\left( \frac{\Phi}{2} \right) ~\sin\left( \frac{\alpha}{2} - \frac{\Phi}{2} \right) \right].
\end{split}
\end{equation}
Table 1 states explicit formulae for $\Psi^\prime_i$ in terms of $I_i$.  Corresponding to Extended Data Figure \ref{fig:inhomogeneous}, one has to construct $\Psi^\prime_i$ according to the correct sextant \cite{hu15}, which is expressed as a set of inequalities of $\Phi$.  In comparison to a homogeneous atmosphere, an inhomogeneous atmosphere has $\omega^\prime$, $x_1$ and $x_2$ as additional parameters.

For an inhomogeneous atmosphere, the reflected light phase curve is still given by equation (\ref{eq:phase_curve}).  The expression relating the secondary eclipse depth $D$ to the geometric albedo $A_g$ is \cite{seager}
\begin{equation}
D = A_g  \left( \frac{R}{a} \right)^2.
\end{equation}
Another way to understand the preceding expression is to realise that the flux of reflected starlight at secondary eclipse is $2 F_0 I_\star$ (the factor of 2 accounts for the fact that the expression for $F$ assumes latitudinal symmetry), which is divided by the flux from a Lambertian disk $\pi I_\star$ to obtain $A_g$.

\renewcommand{\arraystretch}{2.1}
\begin{table*}
\begin{center}
\caption{Formulae for Computing Regions of Enhanced Reflectivity}
\begin{tabular}{ccc}
\hline
\hline
Sextant & Integration Angles & $\Psi^\prime_i$ \\
\hline
$-\frac{\pi}{2} \le \alpha - \frac{\pi}{2} \le \frac{\pi}{2} \le \alpha + x_1 \le \alpha + x_2$ & $\Phi_2 = \frac{\pi}{2}$, $\Phi_1 = \alpha - \frac{\pi}{2}$ & $I_i\left(\Phi_2\right) - I_i\left(\Phi_1\right)$\\
\hline
$-\frac{\pi}{2} \le \alpha - \frac{\pi}{2} \le \alpha + x_1 \le \frac{\pi}{2} \le \alpha + x_2$ & $\Phi_2 = \alpha + x_1$, $\Phi_1 = \alpha - \frac{\pi}{2}$ & $I_i\left(\Phi_2\right) - I_i\left(\Phi_1\right)$\\
\hline
$-\frac{\pi}{2} \le \alpha - \frac{\pi}{2} \le \alpha + x_1 \le \alpha + x_2 \le \frac{\pi}{2}$ & $\Phi_2 = \alpha + x_1$, $\Phi_1 = \alpha - \frac{\pi}{2}$ & $I_i\left(\Phi_2\right) - I_i\left(\Phi_1\right)$ \\
 & $\Phi_4 = \frac{\pi}{2}$, $\Phi_3 = \alpha + x_2$ & $+ I_i\left(\Phi_4\right) - I_i\left(\Phi_3\right)$ \\
\hline
$\alpha + x_1 \le \alpha + x_2 \le -\frac{\pi}{2} \le \alpha + \frac{\pi}{2} \le \frac{\pi}{2} $ & $\Phi_2 = \alpha + \frac{\pi}{2}$, $\Phi_1 = -\frac{\pi}{2}$ & $I_i\left(\Phi_2\right) - I_i\left(\Phi_1\right)$\\
\hline
$\alpha + x_1 \le -\frac{\pi}{2} \le \alpha + x_2 \le \alpha + \frac{\pi}{2} \le \frac{\pi}{2} $ & $\Phi_2 = \alpha + \frac{\pi}{2}$, $\Phi_1 = \alpha + x_2$ & $I_i\left(\Phi_2\right) - I_i\left(\Phi_1\right)$\\
\hline
$-\frac{\pi}{2} \le \alpha + x_1 \le \alpha + x_2 \le \alpha + \frac{\pi}{2} \le \frac{\pi}{2} $ & $\Phi_2 = \alpha + x_1$, $\Phi_1 = -\frac{\pi}{2}$ & $I_i\left(\Phi_2\right) - I_i\left(\Phi_1\right)$  \\
 & $\Phi_4 = \alpha + \frac{\pi}{2}$, $\Phi_3 = \alpha + x_2$  & $+ I_i\left(\Phi_4\right) - I_i\left(\Phi_3\right)$\\
\hline
\hline
\end{tabular}\\
\vspace{0.1in}
Note: The index $i$ refers to the components S, L and C.
\end{center}
\end{table*}

\subsection*{Fit to data}

We simultaneously fit the long-cadence fluxes of the hot Jupiter host Kepler-7 observed by the Kepler mission, spanning the times corresponding to the phase curve and the secondary eclipse. We retrieve the Pre-search Data Conditioning Simple Aperture Photometry (PDCSAP) flux light curve from MAST with \textsf{lightkurve} \cite{Lightkurve2018}, and mask $8\sigma$ outliers in flux. For the purposes of this analysis, we have masked out transits. There are 60,934 remaining fluxes at 30-minute cadence from Quarters 0--17.

The light curve model consists of four components: reflected light, thermal emission, the secondary eclipse and a Gaussian process. All components are fitted simultaneously to the entire out-of-transit light curve.

The reflected light component is modelled with an inhomogeneous atmosphere and the Henyey-Greenstein law of reflection. The planet has two regions defined by the single-scattering albedos $0 \leq \omega_0$ and $\omega = \omega_0+\omega^\prime \leq 1$, with one global scattering asymmetry factor $g$ that follows a Gaussian prior $\mathcal{N}(0, 0.2)$, where the darker surface is bounded by the local longitudes $-\pi/2 < x_1 < x_2 < \pi/2$. We note that the parameter pair $(\omega_0, \omega)$ shares similar constraints to the quadratic limb-darkening parameters \cite{kipping13} and thus adapt the same technique to the single-scattering albedos for efficient, uninformative triangular sampling.

The thermal emission is computed by parameterising the temperature map with a generalised spherical harmonic basis \cite{hw14}. The thermal emission model requires two free parameters: the spherical harmonic power in the temperature map in the $m=\ell=1$ mode $C_{11}$, and the Bond albedo $A_{\rm B}$. The heat redistribution parameter is bounded between 1/4 (full redistribution) and 2/3 (no redistribution), and we select 1/4 for this analysis as it is degenerate with $A_{\rm B}$ \cite{ca11}. Since infrared phase curves have not been measured for Kepler-7 b, an empirical relationship between the peak offset of the thermal component and the irradiation temperature is used \cite{zhang18,beatty19}, which yields an expected peak offset consistent with zero; we fix the orbital phase of maximum thermal emission to $\alpha=0$.  We restrict the retrieved 4.5 $\mu$m secondary eclipse depth of Kepler-7b to be consistent with the previously reported upper limit measured by the Spitzer Space Telescope \cite{demory13}.

The secondary eclipse is modelled with the \textsf{exoplanet} package with no limb darkening \cite{exoplanet:exoplanet,exoplanet:agol20}. The mid-transit time, period, impact parameter, exoplanetary radius and stellar radius \cite{esteves15}, as well as the stellar density \cite{Southworth2012}, are fixed to their previously reported values. The ratio of the exoplanetary flux to the stellar flux is assumed to drop to zero during secondary eclipse.  We account for the 30-minute exposure times by super-sampling the eclipse curve.

Finally, we account for the stellar rotation and time-correlated systematics in the photometry with Gaussian process (GP) regression, using the simple-harmonic oscillator (SHO) kernel implemented in \textsf{celerite2} \cite{celerite2:foremanmackey17,celerite2:foremanmackey18}. We set the kernel undamped period and the damping timescale to the stellar rotation period (15.3 days) and twice the rotation period, respectively, as measured by the autocorrelation function of the observations after transits and eclipses have been masked. This ensures that most of the periodic power in the GP is concentrated near periods similar to the stellar rotation or longer, with considerably less power near the 4.9-day orbital period of the exoplanet. We note that larger-amplitude GP solutions are associated with less severe phase curve asymmetry. 

The reflected light and thermal emission phase curve model components are implemented in \textsf{theano} \cite{exoplanet:theano} for compatibility with the \texttt{exoplanet} software ecosystem for statistical inference with \textsf{PyMC3} \cite{exoplanet:exoplanet,exoplanet:pymc3}.  Metropolis-Hastings sampling is used to infer all of the previously mentioned parameters, as well as the uncertainties on the 30 minute cadence fluxes, and finds a maximum likelihood uncertainty of 130 ppm. The posteriors are integrated until the Gelman-Rubin statistic becomes $\hat{r} \lesssim 1.05$ for all parameters \cite{gr92}.  Several other open-source software packages were used for this research \cite{matplotlib,VanDerWalt2011,exoplanet:astropy13,fm16,exoplanet:astropy18,exoplanet:luger18,vir20}.  We enumerate the maximum-likelihood parameters and other derived quantities in Table~\ref{tab:pcparams}.

\subsection*{Re-parameterisation}

For the Henyey-Greenstein reflection law, one may treat $\omega_0$, $\omega^\prime$, $x_1$, $x_2$ and $A_g$ as the independent parameters and express $g$ as a dependent parameter,
\begin{equation}
g = \frac{-\left(2C_3+1\right) \pm \sqrt{1 + 8C_3}}{2C_3},
\end{equation}
where the various quantities are
\begin{equation}
\begin{split}
C_3 =& \frac{ 16 \pi A_g - 32 C_1 - 2 \pi \omega_0 C - \pi \omega^\prime C_2 C^\prime }{2 \pi \omega_0 + \pi \omega^\prime C_2}, \\
C_2 =& 2 + \sin x_1 - \sin x_2, \\
C_1 =& \frac{\omega_0 \rho_{\rm L_0} \pi}{12} + \frac{\omega^\prime \rho^\prime_{\rm L}}{12} \left[ x_1 - x_2 + \pi \right. \\
&\left. + \frac{1}{2} \left( \sin{2x_1} - \sin{2x_2} \right) \right] + \frac{\pi \omega_0 \rho_{\rm C_0}}{32} \\
&+ \frac{3\pi \omega^\prime \rho^\prime_{\rm C}}{64} \left[ \frac{2}{3} + \frac{3}{8}\left( \sin x_1 - \sin x_2 \right) \right.\\
&\left. + \frac{1}{24} \left( \sin{3x_1} - \sin{3x_2} \right) \right], \\
C =& -1 + \frac{1}{4}\left( 1 + \epsilon \right)^2 \left( 2 - \epsilon \right)^2, \\
C^\prime =& -1 + \frac{1}{4}\left( 1 + \epsilon^\prime \right)^2 \left( 2 - \epsilon^\prime \right)^2, \\
\epsilon^\prime =& \frac{1-\gamma^\prime}{1+\gamma^\prime}, \\
\gamma^\prime =& \sqrt{1-\omega^\prime}.
\end{split}
\end{equation}

\begin{table}[!t]
\centering
\begin{tabular}{lr}
\hline
\hline
Parameter & Value \\ 
\hline 
Single-scattering albedo, $\omega_0$ & ${0.0136}^{+0.0132}_{-0.0093}$ \\
Single-scattering albedo, $\omega^\prime$ & ${0.115}^{+0.044}_{-0.049}$ \\
Scattering asymmetry factor, $g$ & ${0.07}^{+0.12}_{-0.11}$ \\
Western limit, $x_1$ (degree) & ${-6.0}^{+4.4}_{-5.0}$ \\
Eastern limit, $x_2$ (degree)& ${40.0}^{+3.3}_{-3.4}$ \\
Spherical harmonic power, $C_{11}$ & ${0.04993}^{+0.11807}_{-0.00014}$ \\
Bond albedo, $A_{\rm B}$ & ${0.20}^{+0.19}_{-0.15}$ \\
Geometric albedo, $A_g$ & ${0.251}^{+0.012}_{-0.016}$ \\
Phase integral, $q$ & ${1.774}^{+0.068}_{-0.073}$ \\
Spherical albedo, $A_s$ & ${0.442}^{+0.023}_{-0.028}$ \\
\hline
\hline
\end{tabular}
\caption{Maximum-likelihood parameters inferred from fitting the Kepler light curve of Kepler-7b with a model consisting of the inhomogeneous atmosphere in reflected light, thermal emission, a secondary eclipse and a Gaussian process. 1$\sigma$ uncertainties are stated. \label{tab:pcparams}}
\end{table}


\begin{thebibliography}{99}

\scriptsize

\bibitem{bond1861} Bond, G.P. On the Light of the Sun, Moon, Jupiter, and Venus. \textit{Mon. Not. R. Astron. Soc.} \textbf{21}, 197--202 (1861).

\bibitem{russell1916} Russell, H.N. On the Albedo of the Planets and their Satellites. \textit{Astrophys. J.} \textbf{43}, 173--196 (1916).

\bibitem{horak50} Horak, H.G. Diffuse Reflection by Planetary Atmospheres. \textit{Astrophys. J.} \textbf{112}, 445--463 (1950).

\bibitem{vau64} de Vaucouleurs, G. Geometric and photometric parameters of the terrestrial planets. \textit{Icarus} \textbf{3}, 187--235 (1964).

\bibitem{sobolev} Sobolev, V.V. Light Scattering in Planetary Atmospheres. Oxford: Pergamon Press (1975).

\bibitem{seager} Seager, S. Exoplanet Atmospheres. Princeton: Princeton University Press (2010).

\bibitem{horak65} Horak, H.G. \& Little, S.J. Calculations of Planetary Reflection. \textit{Astrophys. J. Suppl.} \textbf{11}, 373--428 (1965).

\bibitem{dy74} Dlugach, J. \& Yanovitskij, E.G. The Optical Properties of Venus and the Jovian Planets. II. Methods and Results of Calculations of the Intensity of Radiation Diffusely Reflected from Semi-infinite Homogeneous Atmospheres. \textit{Icarus} \textbf{22}, 66--81 (1974).

\bibitem{hh89} Hovenier, J.W. \& Hage, J.I. Relations involving the spherical albedo and other photometric quantities of planets with thick atmospheres. \textit{Astron. \& Astrophys.} \textbf{214}, 391--401 (1989).

\bibitem{marley99} Marley, M.S., Gelino, C., Stephens, D., Lunine, J.I. \& Freedman, R. Reflected Spectra and Albedos of Extrasolar Giant Planets. I. Clear and Cloudy Atmospheres. \textit{Astrophys. J.} \textbf{513}, 879--893 (1999). 

\bibitem{sudarsky00} Sudarsky, D., Burrows, A., \& Pinto, P. Albedo and Reflection Spectra of Extrasolar Giant Planets. \textit{Astrophys. J.} \textbf{538}, 885--903 (2000). 

\bibitem{van74} van de Hulst, H.C. The Spherical Albedo of a Planet Covered with a Homogeneous Cloud Layer. \textit{Astron. \& Astrophys.} \textbf{35}, 209--214 (1974).

\bibitem{madhu12} Madhusudhan, N. \& Burrows, A. Analytic Models for Albedos, Phase Curves and Polarization of Reflected Light from Exoplanets. \textit{Astrophys. J.} \textbf{747}, 25 (2012).

\bibitem{hapke81} Hapke, B. Bidirectional Reflectance Spectroscopy. 1. Theory. \textit{J. Geophys. Res.} \textbf{86}, 3039--3054 (1981).

\bibitem{chandra} Chandrasekhar, S. Radiative Transfer. New York: Dover Publications (1960).

\bibitem{pierrehumbert} Pierrehumbert, R.T. Principles of Planetary Climate. Cambridge: Cambridge University Press (2010).

\bibitem{demory13} Demory, B.-O. Inference of Inhomogeneous Clouds in an Exoplanet Atmosphere. \textit{Astrophys. J. Lett.} \textbf{776}, L25 (2013).

\bibitem{hu15} Hu, R., Demory, B.-O., Seager, S., Lewis, N. \& Showman, A.P. A Semi-analytical Model of Visible-wavelength Phase Curves of Exoplanets and Applications to Kepler-7b and Kepler-10b. \textit{Astrophys. J.} \textbf{802}, 51 (2015).

\bibitem{sh15} Shporer, A. \& Hu, R. Studying Atmosphere-dominated Hot Jupiter Kepler Phase Curves: Evidence that Inhomogeneous Atmospheric Reflection Is Common. \textit{Astronom. J.} \textbf{150}, 112 (2015).

\bibitem{oreshenko16} Oreshenko, M., Heng, K. \& Demory, B.-O. Optical phase curves as diagnostics for aerosol composition in exoplanetary atmospheres. \textit{Mon. Not. R. Astron. Soc.} \textbf{457}, 3420--3429 (2016).

\bibitem{esteves15} Esteves, L.J., De Mooij, E.J.W. \& Jayawardhana, R. Changing Phases of Alien Worlds: Probing Atmospheres of Kepler Planets with High-precision Photometry. \textit{Astrophys. J.} \textbf{804}, 150 (2015).

\bibitem{kipping10} Kipping, D.M. Binning is sinning: morphological light-curve distortions due to finite integration time. \textit{Mon. Not. R. Astron. Soc.} \textbf{408}, 1758--1769 (2010).

\bibitem{burrows06} Burrows, A., Sudarsky, D. \& Hubeny, I. L and T Dwarf Models and the L to T Transition. \textit{Astrophys. J.} \textbf{640}, 1063–-1077 (2006).

\bibitem{hl21} Heng, K., \& Li, L. Jupiter as an Exoplanet: Insights from Cassini Phase Curves. \textit{Astrophys. J. Lett.}, \textbf{909}, L20 (2021).

\bibitem{sc15} Schwartz, J.C. \& Cowan, N.B. Balancing the energy budget of short-period giant planets: evidence for reflective clouds and optical absorbers. \textit{Mon. Not. R. Astron. Soc.} \textbf{449}, 4192--4203 (2015).

\bibitem{mihalas} Mihalas, D. Stellar Atmospheres. San Francisco: W.H. Freeman and Company (1970).

\bibitem{toon89} Toon, O.B., McKay, C.P., Ackerman, T.P. \& Santhanam, K. Rapid Calculation of Radiative Heating Rates and Photodissociation Rates in Inhomogeneous Multiple Scattering Atmospheres. \textit{J. Geophys. Res.} \textbf{94}, 16287--16301 (1989).

\bibitem{heng} Heng, K. Exoplanetary Atmospheres. Princeton: Princeton University Press (2017).

\bibitem{heng18} Heng, K., Malik, M. \& Kitzmann, D. Analytical Models of Exoplanetary Atmospheres. VI. Full Solutions for Improved Two-stream Radiative Transfer, Including Direct Stellar Beam. \textit{Astrophys. J. Suppl.} \textbf{237}, 29 (2018).

\bibitem{lommel} Lommel, E. Die Photometrie der diffusen Zur\"{u}ckwerfung. \textit{Bayerische Akademie der Wissenschaften M\"{u}nchen Sitzungsberichte}, \textbf{17}, 95--132 (1887).

\bibitem{seeliger} Seeliger, H. Zur Photometrie zerstreut reflectirender Substanzen. \textit{Bayerische Akademie der Wissenschaften M\"{u}nchen Sitzungsberichte}, \textbf{18}, 201--248 (1888).

\bibitem{davidovic08} Davidovi\'{c}, D.M., Vukani\'{c}, J., \& Arsenovi\'{c}, D. Two new analytic approximations of the Chandrasekhar's H function for isotropic scattering. \textit{Icarus} \textbf{194}, 389--397 (2008).

\bibitem{hg} Henyey, L.G. \& Greenstein, J.L. Diffusion radiation in the Galaxy. \textit{Astrophys. J.} \textbf{93}, 70--83 (1941).

\bibitem{li18} Li, L., et al. Less absorbed solar energy and more internal heat for Jupiter. \textit{Nat. Commun.} \textbf{9}, 3709 (2018).

\bibitem{Lightkurve2018} Lightkurve Collaboration. Lightkurve: Kepler and TESS time series analysis in Python. \textit{Astrophysics Source Code Library} http://ascl.net/1812.013 (2018).

\bibitem{kipping13} Kipping, D.M. Efficient, uninformative sampling of limb darkening coefficients for two-parameter laws. \textit{Mon. Not. R. Astron. Soc.} \textbf{435}, 2152--2160 (2013).

\bibitem{hw14} Heng, K. \& Workman, J. Analytical Models of Exoplanetary Atmospheres. I. Atmospheric Dynamics via the Shallow Water System. \textit{Astrophys. J. Suppl. Ser.} \textbf{213}, 27 (2014).

\bibitem{ca11} Cowan, N.B. \& Agol, E. The Statistics of Albedo and Heat Redistribution on Hot Exoplanets. \textit{Astrophys. J.} \textbf{729}, 54 (2011).

\bibitem{zhang18} Zhang, M. et al. Phase Curves of WASP-33b and HD 149026b and a New Correlation between Phase Curve Offset and Irradiation Temperature. \textit{Astronom. J.} \textbf{155}, 83 (2018).

\bibitem{beatty19} Beatty, T.G.,et al. Spitzer Phase Curves of KELT-1b and the Signatures of Nightside Clouds in Thermal Phase Observations. \textit{Astronom. J.} \textbf{158}, 66 (2019).

\bibitem{exoplanet:exoplanet} Foreman-Mackey, D., Czekala, I., Agol, E., Luger, R. \& Barclay, T. dfm/exoplanet: exoplanet v0.2.4. \textit{Zenodo} https://doi.org/10.5281/zenodo.3595344 (2019).

\bibitem{exoplanet:agol20} Agol, E., Luger, R. \& Foreman-Mackey, D. Analytic Planetary Transit Light Curves and Derivatives for Stars with Polynomial Limb Darkening. \textit{Astronom. J.} \textbf{159}, 123 (2020).

\bibitem{Southworth2012} Southworth, J. Homogeneous studies of transiting extrasolar planets - V. New results for 38 planets. \textit{Mon. Not. R. Astron. Soc.} \textbf{426}, 1291--1323 (2012).

\bibitem{celerite2:foremanmackey17} Foreman-Mackey, D., Agol, E., Ambikasaran, S. \& Angus, R. Fast and Scalable Gaussian Process Modeling with Applications to Astronomical Time Series. \textit{Astronom. J.} \textbf{154}, 220 (2017).

\bibitem{celerite2:foremanmackey18} Foreman-Mackey, D. Scalable Backpropagation for Gaussian Processes using Celerite. \textit{Res. Notes AAS} \textbf{2}, 31 (2018).

\bibitem{exoplanet:theano} The Theano Development Team. Theano: A Python framework for fast computation of mathematical expressions. Preprint at https://arxiv.org/abs/1605.02688 (2016).

\bibitem{exoplanet:pymc3} Salvatier, J., Wiecki, T.V. \& Fonnesbeck, C. PyMC3: Python probabilistic programming framework. https://doi.org/10.7717/peerj-cs.55 (2016).

\bibitem{gr92} Gelman, A. \& Rubin, D.B. Inference from Iterative Simulation Using Multiple Sequences. \textit{Stat. Sci.} \textbf{7}, 457--472 (1992).

\bibitem{matplotlib} Hunter, J.D. Matplotlib: A 2D Graphics Environment. \textit{Comput. Sci. Eng.} \textbf{9}, 90--95 (2007).

\bibitem{VanDerWalt2011} van der Walt, S., Colbert, S.C. \& Varoquaux, G. The NumPy Array: A Structure for Efficient Numerical Computation. \textit{Comput. Sci. Eng.} \textbf{13}, 22--30 (2011).

\bibitem{exoplanet:astropy13} Astropy Collaboration. Astropy: A community Python package for astronomy. \textit{Astron. \& Astrophys.} \textbf{558}, A33 (2013).

\bibitem{fm16} Foreman-Mackey, D. corner.py: Scatterplot matrices in Python. \textit{J. Open Source Softw.} \textbf{1}, 24 (2016).

\bibitem{exoplanet:astropy18} Astropy Collaboration. The Astropy Project: Building an Open-science Project and Status of the v2.0 Core Package. \textit{Astronom. J.} \textbf{156}, 123 (2018).

\bibitem{exoplanet:luger18} Luger, R. et al. starry: Analytic Occultation Light Curves. \textit{Astronom. J.} \textbf{157}, 64 (2019).

\bibitem{vir20} Virtanen, P. et al. SciPy 1.0: fundamental algorithms for scientific computing in Python. \textit{Nat. Methods} \textbf{17}, 261--272 (2020).

\end{thebibliography}
\end{document}